\documentclass[aps,prb,twocolumn,superscriptaddress,floatfix]{revtex4-1}
\usepackage{graphicx}
\usepackage{amssymb}
\usepackage{amsmath}
\usepackage{upgreek}
\usepackage{dsfont}
\usepackage{color}
\newcommand{\vect}[1]{\boldsymbol{\mathbf{#1}}}       
\newcommand{\mathcomma}{~ ,}            
\newcommand{\mathperiod}{~ .}           
\newcommand{\dosf}{\rho_{\mathrm{F}}}     

\DeclareMathOperator{\sgn}{sgn}

\renewcommand{\Re}{\operatorname{\mathrm{Re}}}
\renewcommand{\Im}{\operatorname{\mathrm{Im}}}
\newcommand{\hpsi}{\hat\psi}
\newcommand{\hPsi}{\hat\Psi}
\newcommand{\bk}{{\bf k}}
\newcommand{\bK}{{\bf K}}
\newcommand{\bnil}{{\bf0}}
\newcommand{\bQ}{{\bf Q}}
\newcommand{\bR}{{\bf R}}
\newcommand{\hU}{\hat U}
\newcommand{\hK}{\hat K}
\newcommand{\hT}{\hat T}
\begin{document}
\title{Pair breaking due to orbital magnetism in iron-based superconductors}
\author{M.\ Hoyer}
\affiliation{Institut f\"ur Theorie der Kondensierten
Materie, Karlsruher Institut f\"ur Technologie, D-76131 Karlsruhe, Germany}
\affiliation{Institut f\"ur Festk\"orperphysik, Karlsruher Institut f\"ur Technologie, D-76021 Karlsruhe, Germany}
\author{M.\,S.\ Scheurer}
\affiliation{Institut f\"ur Theorie der Kondensierten
Materie, Karlsruher Institut f\"ur Technologie, D-76131 Karlsruhe, Germany}
\author{S.\,V.\ Syzranov}
\affiliation{Department of Physics, University of Colorado, Boulder, CO 80309, USA}
\affiliation{Institut f\"ur Theorie der Kondensierten
Materie, Karlsruher Institut f\"ur Technologie, D-76131 Karlsruhe, Germany}
\author{J.\ Schmalian}
\affiliation{Institut f\"ur Theorie der Kondensierten
Materie, Karlsruher Institut f\"ur Technologie, D-76131 Karlsruhe, Germany}
\affiliation{Institut f\"ur Festk\"orperphysik, Karlsruher Institut f\"ur Technologie, D-76021 Karlsruhe, Germany}
\date{\today}
\begin{abstract}
We consider superconductivity in the presence of impurities in a two-band model suited for the description of iron-based superconductors. 
We analyze the effect of interband scattering processes on superconductivity, allowing for orbital, i.\,e., nonspin-magnetic but time-reversal symmetry-breaking impurities. 
Pair breaking in such systems is described by a nontrivial phase in an interband-scattering matrix element. 
We find that the transition temperature of conventional superconductors can be suppressed due to interband scattering, whereas unconventional superconductors may be unaffected. 
We also discuss the stability of density wave phases in the presence of impurities. 
As an example, we consider impurities associated with imaginary charge density waves that are of interest for iron-based superconductors.  
\end{abstract}
\maketitle
\section{Introduction}
\begin{figure}[b]
 \includegraphics[width=0.75\columnwidth]{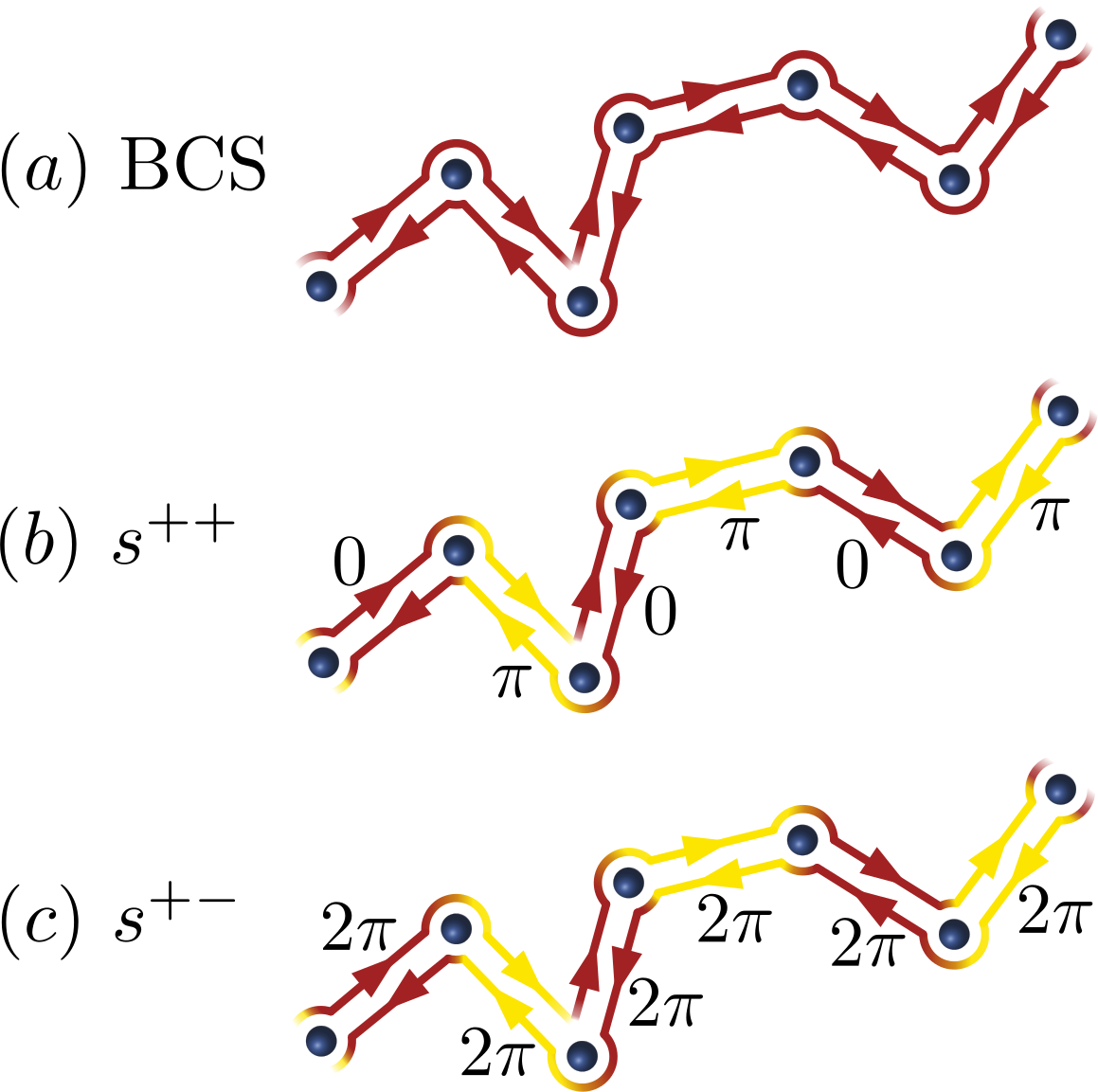}
 \caption{(Color online) Visualization of (a) the Anderson theorem in single-band $s$-wave superconductors, (b) pair breaking as a consequence of interband scattering due to time-reversal symmetry-breaking impurities in a two-band $s^{++}$ superconductor, and (c) an analog of the Anderson theorem for the two-band $s^{+-}$ superconductor. The broken lines correspond to the trajectories of electrons scattered by impurities, where different colors indicate electrons carrying different band indices.}
 \label{fig:visualization}
\end{figure}
Conventional superconductivity is astonishingly robust against impurity scattering. The transition temperature~$T_\mathrm{c}$ remains approximately constant in the presence of nonmagnetic impurities as follows from the Anderson theorem\cite{JPhysChemSolids.11.26,JETP.8.1090,JETP.9.220}. 
The physical reason for this protection against nonmagnetic impurities is visualized in Fig.~\ref{fig:visualization}(a). Superconductivity occurs as a consequence of an effective attraction between electrons which is mediated by the distorted lattice, and this coupling is largest for electrons on time-reversed paths. In a disordered material, the trajectories are changed due to scattering, but the coupling remains unaffected as long as the disorder strength is weak. 
Magnetic impurities, on the other hand, are pair breaking for conventional superconductors and suppress the transition temperature\cite{Soviet.Phys.JETP.12.1243} that vanishes at a critical value of the scattering rate. 
Unconventional superconductors, in contrast, are already sensitive to nonmagnetic impurities\cite{Golubov.Mazin.1997,Kogan2009}, and again, superconductivity vanishes at a critical scattering rate. The suppression of $T_\mathrm{c}$ with increasing scattering rate due to nonmagnetic impurities is therefore considered as a signature of unconventional superconductivity. 

In iron-based superconductors, there is strong evidence supporting an $s^{+-}$ scenario for superconductivity in these materials, where the pairing gap changes sign between different bands without breaking a point group symmetry. However, the pairing state is still under debate\cite{HirschfeldMazin2011,Chubukov2012-review}, and in particular the  relatively weak suppression of the superconducting transition temperature with increasing concentration of nonmagnetic impurities has been used as an argument in favor of a conventional pairing state\cite{PhysRevB.84.020513,Onari.Kontani.2009.ViolationAnderson}. 
One explanation for this behavior is that intraband and interband scattering are not equally strong in iron-based superconductors, and transport properties are mainly determined by intraband scattering effects, whereas the suppression of $T_\mathrm{c}$ is due to interband scattering\cite{Vorontsov.2009}. 
Moreover, in this paper, we show that the discrimination of $s^{++}$ and $s^{+-}$~pairing state based on their response to the presence of apparently nonmagnetic impurities is not always possible. 

The iron-based systems are multiband superconductors in which electrons from different orbitals contribute to superconductivity and/or magnetic order. Furthermore, competing states of order are a characteristic of these materials. Model calculations\cite{Chubukov2008,Podolsky2009,KangTesanovic2011} show that imaginary charge density waves are expected to compete with antiferromagnetism and superconductivity in these materials. Such imaginary charge density wave order could, similar to spin density wave order\cite{Millis.Morr.Schmalian.2001,GastiasoroAndersen2013,Gastiasoro2013-Dimers}, nucleate around nonmagnetic impurities which then break time-reversal symmetry and can  thereby be associated with (orbital) magnetism. Thus, the detailed impact of orbital-magnetic impurities on pairing in these multiband systems is an interesting open 
topic. 

In this paper we consider a two-band model for iron-based superconductors with impurities causing intraband and interband scattering processes. In contrast to previous studies\cite{Onari.Kontani.2009.ViolationAnderson,SengaKontani2008,EfremovHirschfeld2011,VavilovChubukov2011,EfremovGolubovDolgov2013,DolgovGolubov2013,HoyerSchmalian2014phasecompetition} which all concentrated on either nonmagnetic or spin-magnetic impurities in iron-based superconductors, we investigate how the interplay between pairing and orbital magnetism takes place. 
In particular, we find that impurities associated with orbital magnetism can lead to the
suppression of $T_\mathrm{c}$ in conventional superconductors which is visualized in Fig.~\ref{fig:visualization}(b). Scattering on such impurities involves a relative phase of~$\pi$, and therefore the interaction matrix element between the electrons accumulates a random phase factor (which is a multiple of $\pi$) that destroys superconductivity if the mean-free path becomes smaller than the superconducting coherence length. 
In addition, we will see that the transition temperature in unconventional superconductors may remain unaffected, i.\,e., there exists an Anderson theorem for the $s^{+-}$~pairing state which is protected against time-reversal symmetry-breaking interband scattering. This protection is a result of an additional phase of~$\pi$ associated with the coupling matrix element which results in a total phase of~$2\pi$ associated with every interband scattering process which is of no importance. This is sketched in Fig.~\ref{fig:visualization}(c). 
As we will show, these effects can be due to impurities that nucleate local orbital-magnetic states. Therefore, it is important for our theory that we allow for spatially extended impurity potentials. 
\section{Disordered two-band model}
We consider a two-band superconductor with impurities, described by the Hamiltonian
$\mathcal{\hat H}=\mathcal{\hat H}_0+\mathcal{\hat H}_\mathrm{int}+\mathcal{\hat H}_\mathrm{dis}$. 
The noninteracting part is given by 
\begin{equation}
      \mathcal{\hat H}_0= \sum_{\vect{k},\sigma}\sum_\alpha\xi_{\alpha,\vect{k}}
      \hpsi_{\alpha,\vect{k},\sigma}^\dagger\hpsi_{\alpha,\vect{k},\sigma}^{}\mathcomma
      \label{eq:cleanhamiltonian}
\end{equation}
where $\alpha$ labels the two bands, $\sigma$ denotes spin, and $\xi_{\alpha,\vect{k}}=\varepsilon_{\alpha,\vect{k}}-\mu$ is the dispersion of band~$\alpha$, measured from the chemical potential. 
We assume that the quasiparticles in band~1 have small momenta near the center
of the Brillouin zone ($\Gamma$ point), while the momenta of quasiparticles in band~2 are close to $\bQ$,
where $2\bQ$ is a reciprocal primitive vector, as it is suitable for iron-based superconductors.
The concrete form of the dispersion relation is not important for our calculations as long as $\xi_{\alpha,\vect{k}}=\xi_{\alpha,-\vect{k}}$ holds. 
For simplicity we assume the density of states near the Fermi level to have the same value~$\rho_\mathrm{F}$ in both bands. The generalization to different densities of states in the two bands is straightforward. Note that, as a consequence of assuming a constant density of states at the Fermi level, our results are independent of the shape of the two Fermi surfaces. In particular, the possibility of an elliptical electron Fermi surface, which is important in the context of magnetic order in iron-based superconductors, is thereby included in our considerations. 

Furthermore, we consider superconductivity (SC) due to interband pairing, described in a BCS-like model, 
\begin{align}\begin{split}
  \mathcal{\hat H}_\mathrm{int}&=\sum_{\vect{k},\vect{k}^\prime}\sum_\alpha V^{\alpha\bar{\alpha}}_{\vect{k},\vect{k}^\prime}\hpsi_{\alpha,\vect{k},\uparrow}^\dagger\hpsi_{\alpha,-\vect{k},\downarrow}^\dagger\hpsi_{\bar{\alpha},-\vect{k}^\prime,\downarrow}^{}\hpsi_{\bar{\alpha},\vect{k}^\prime,\uparrow}^{} \mathcomma
  \label{eq:hamiltonian-int}\\
  V^{\alpha\bar{\alpha}}_{\vect{k},\vect{k}^\prime}&=\left\{\begin{matrix} V & \text{for }|\xi_{\alpha,\vect{k}}|,|\xi_{\bar{\alpha},\vect{k}^\prime}|<\Lambda\mathcomma \\ 0 & \text{otherwise}\mathcomma \end{matrix} \right. 
\end{split}
\end{align}
where $\bar{\alpha}$ labels the band other than $\alpha$. 

The most generic Hamiltonian of disorder in such a system reads 
\begin{align}
  \hat{\mathcal{H}}_\mathrm{dis}=\sum_{\alpha,\beta}\sum_{s,s^\prime}\hpsi^\dagger_\alpha(\vect{R}_s)W_{\alpha\beta}(\vect{R}_s,\vect{R}_{s^\prime})\hpsi_\beta(\vect{R}_{s^\prime})\mathcomma \label{eq:disorderhamiltonian1}
\end{align}
where the indices $s$ and $s^\prime$ label lattice sites $\vect{R}_s$ and $\vect{R}_{s^\prime}$. 
Here the $\hpsi_\alpha(\vect{R}_s)$ and $W_{\alpha\beta}(\vect{R}_s,\vect{R}_{s^\prime})$ are vectors and matrices in spin space, respectively, i.\,e., $\hpsi_{\alpha}(\vect{R}_s)=(\begin{matrix} \hpsi_{\alpha,\uparrow}(\vect{R}_s)\mathcomma & \hpsi_{\alpha,\downarrow}(\vect{R}_s)\end{matrix})^T$, 
where by $\hpsi^{(\dagger)}_{\alpha,\sigma}(\vect{R})$ we denote field operators in position space which have to be understood as convolution with momenta in band~$\alpha$ only.  

This disorder is typically represented by identical impurities with random locations~$\vect{R}_i$,  
\begin{align}
\begin{split}
  \mathcal{\hat H}_\mathrm{dis}&=\sum_{i=1}^N\hU_{\vect{R}_i}\mathcomma \\ 
  \hU_{\vect{R}_i}&= \sum_{\alpha,\beta}\sum_{s,s^\prime}\hpsi_{\alpha}^\dagger(\vect{R}_s+\vect{R}_i)J_{ss^\prime}^{\alpha\beta}\hpsi^{}_{\beta}(\vect{R}_{s^\prime}+\vect{R}_i)\mathperiod \label{eq:disorderhamiltonian2}
\end{split}
\end{align}
These two formulations of the impurity Hamiltonian, Eqs.~\eqref{eq:disorderhamiltonian1} and \eqref{eq:disorderhamiltonian2}, are connected by 
\begin{equation}
  W_{\alpha\beta}(\vect{R}_s,\vect{R}_{s^\prime})=\sum_iJ^{\alpha\beta}_{s-i,s^\prime-i} \mathperiod
\end{equation}
The matrix element $J_{ss^\prime}^{\alpha\beta}$ can account for intraband ($\alpha=\beta$) as well as interband ($\alpha\neq\beta$) scattering processes. In general, $J_{ss^\prime}^{\alpha\beta}$ are the matrix elements of  a nondiagonal matrix in position space, allowing us to describe spatially extended scattering centers, which is essential, e.\,g., to account for
orbital-magnetic impurities.

At the same time, in what follows, we assume for simplicity that the disorder is short correlated
on the scale $k_\mathrm{F}^{-1}$, where $k_\mathrm{F}$ is the largest of the Fermi wave vectors in the two bands. 
\section{Symmetry considerations}\label{sec:symmetryanalysis}
Before we explicitly calculate the effect of impurities on the SC~transition temperature of $s^{++}$ and $s^{+-}$~superconductors, we will provide an extension of Anderson's theorem\cite{JPhysChemSolids.11.26,JETP.8.1090,JETP.9.220} for two-band superconductors. Specifically, it will be demonstrated that the $s^{++}$~pairing state is robust against time-reversal-symmetric (TRS) scattering while for the $s^{+-}$~pairing state, the gap is unchanged by time-reversal-antisymmetric (TRA) interband and TRS intraband disorder. Furthermore, we present a criterion for the protection of density waves in the presence of disorder.

We consider the two-band $s$-wave superconductor as defined in Eqs.~\eqref{eq:cleanhamiltonian} and \eqref{eq:hamiltonian-int}. The corresponding mean-field Hamiltonian is given by 
\begin{align}
\begin{split}
 &\hat{\mathcal{H}}^{\mathrm{MF}}_\mathrm{SC} = \sum_{\vect{k},\alpha} \hpsi_{\alpha,\vect{k}}^\dagger \xi_{\alpha,\vect{k}}\hpsi_{\alpha,\vect{k}}^{} \\
  & \quad+ \sum_{\vect{k},\alpha}\frac{\Delta_\alpha}{2} \Big[\hpsi_{\alpha,\vect{k}}^\dagger \mathrm{i}\hat{\sigma}_2(\hpsi^\dagger_{\alpha,-\vect{k}})^T + \hpsi^T_{\alpha,-\vect{k}}\left(\mathrm{i}\hat{\sigma}_2\right)^\dagger\hat{\psi}_{\alpha,\vect{k}}^{}\Big] \mathcomma
 \label{MFCleanHam}
\end{split}
\end{align} 
where $\Delta_{\alpha}\in\mathds{R}$ denotes the  pairing in band $\alpha$ which is taken to be momentum independent ($s$~wave), as in Eq.~\eqref{eq:hamiltonian-int}. 
The mean-field Hamiltonian~\eqref{MFCleanHam} with a homogeneous order parameter~$\Delta_\alpha$ can be applied to a disordered system provided the disorder strength is sufficiently weak\cite{Bulaevskii1984,Sadovskii1997}, so that electron states near the Fermi surface are delocalized in the normal metal phase ($\Delta_\alpha=0$) or the localization length is large,
\begin{equation}
\xi\gg(T_\mathrm{c}\rho_\mathrm{F})^{-\frac{1}{d}},
\label{eq:localizationlengthcriterion}
\end{equation}
where $T_\mathrm{c}$ is the critical temperature of the superconductive transition. 
Throughout the paper we assume that $k_\mathrm{F}l\gg1$, where $l$ is the mean-free path close to the Fermi surface in the normal phase. This condition, in particular, ensures the absence of localization in 3D~materials and, thus, the applicability of the mean-field Hamiltonian~\eqref{MFCleanHam}. In the case of a 2D~material, we assume additionally that the condition \eqref{eq:localizationlengthcriterion} is fulfilled. 

We introduce Nambu spinors $ \hPsi_{\alpha}(\vect{k}) = (\begin{matrix} \hpsi_{\alpha,\vect{k}}^{}\mathcomma & \mathrm{i} \hat{\sigma}_2 (\hpsi^\dagger_{\alpha,-\vect{k}})^T \end{matrix})^T$ and $\hPsi^\dagger_{\alpha}(\vect{k}) = (\begin{matrix} \hat{\psi}_{\alpha,\vect{k}}^\dagger\mathcomma & \hat{\psi}^T_{\alpha,-\vect{k}} \left( \mathrm{i}\hat{\sigma}_2\right)^\dagger \end{matrix})$ 
to write the mean-field Hamiltonian~\eqref{MFCleanHam} in the quadratic form
\begin{equation}
  \hat{\mathcal{H}}^{\mathrm{MF}}_\mathrm{SC} = \frac{1}{2}\sum_{\vect{k},\alpha}\hPsi_\alpha^\dagger(\vect{k}) \begin{pmatrix} \xi_{\alpha,\vect{k}} & \Delta_\alpha \\ \Delta_\alpha & -\xi_{\alpha,\vect{k}} \end{pmatrix} \hPsi_\alpha(\vect{k})\mathperiod
\end{equation} 
It is convenient to consider a given disorder realization\cite{Potter2011}, as described by the general quadratic term~\eqref{eq:disorderhamiltonian1}, in momentum space, where it reads 
\begin{widetext}
\begin{align}
 \hat{\mathcal{H}}_{\text{dis}} &= \frac{1}{2} \sum_{\vect{k},\vect{k}^\prime}\sum_{\alpha,\alpha^\prime} \hPsi_{\alpha}^\dagger(\vect{k}) \begin{pmatrix} W_{\alpha,\alpha'}(\vect{k},\vect{k}') & 0 \\ 0 & -i\hat{\sigma}_2 W^T_{\alpha',\alpha}(-\vect{k}',-\vect{k}) (i\hat{\sigma}_2)^\dagger \end{pmatrix} \hPsi_{\alpha'}(\vect{k}') \mathperiod
\end{align} 
\end{widetext}
The only constraint on $W_{\alpha,\alpha'}(\vect{k},\vect{k}')$ is $W^\dagger_{\alpha',\alpha}(\vect{k}',\vect{k}) = W_{\alpha,\alpha'}(\vect{k},\vect{k}') $
due to Hermiticity. For the following analysis of time-reversal symmetry, it is convenient to split $W_{\alpha,\alpha'}(\vect{k},\vect{k}')$ according to
\begin{equation}
 W_{\alpha,\alpha^\prime}(\vect{k},\vect{k}^\prime) = W_{\alpha,\alpha^\prime}^+(\vect{k},\vect{k}^\prime) + W_{\alpha,\alpha^\prime}^-(\vect{k},\vect{k}^\prime)
\end{equation}
into parts that are symmetric and antisymmetric under time reversal, 
\begin{equation}
 W_{\alpha,\alpha'}^{\pm}(\vect{k},\vect{k}') \equiv \frac{1}{2} \big[W_{\alpha,\alpha'}(\vect{k},\vect{k}') \pm \hat{T} W_{\alpha,\alpha'}(-\vect{k},-\vect{k}') \hat{T}^{-1} \big],
\end{equation} 
where $\hat{T} = \mathrm{i}\hat{\sigma}_2\hat{\mathcal{K}}$ denotes the time-reversal operator for spin-$\tfrac{1}{2}$, with $\hat{\mathcal{K}}$ representing complex conjugation. 
Introducing Pauli matrices $\hat{\tau}_i$ acting in band space, and defining $\Delta_{\pm}=\frac{1}{\sqrt{2}}(\Delta_1\pm\Delta_2)$, the Hamiltonian can be written compactly as $\hat{\mathcal{H}}^\mathrm{MF}_\mathrm{SC}+\hat{\mathcal{H}}_\mathrm{dis}=\frac{1}{2}\sum_{\vect{k},\vect{k}^\prime}\sum_{\alpha,\alpha^\prime}\hPsi_\alpha^\dagger(\vect{k})\hat{h}_{\alpha\alpha^\prime}(\vect{k},\vect{k}^\prime)\hPsi_{\alpha^\prime}(\vect{k}^\prime)$ 
with 
\begin{equation}
 \hat{h} = \begin{pmatrix} \hat{\xi} + \hat{W}^+ + \hat{W}^- & \frac{1}{\sqrt{2}}\left(\Delta_+\hat{\tau}_0 + \Delta_-\hat{\tau}_3\right) \\ \frac{1}{\sqrt{2}}\left(\Delta_+\hat{\tau}_0 + \Delta_-\hat{\tau}_3\right) & -\left(\hat{\xi} + \hat{W}^+ - \hat{W}^-\right) \end{pmatrix}\mathcomma \label{DisorderedBdGHam}
\end{equation}
where $\hat{\xi}$ is the diagonal matrix of band energies $\xi_\alpha(\vect{k})$. 

The spectrum of $\hat{h}$ is found by solving $\det(\hat{h}-\epsilon\hat{\mathds{1}})=0$ for $\epsilon$, where we can use that 
\begin{equation}
 \det\begin{pmatrix}A& B\\ C&D\end{pmatrix}=\det\left(AD-CB\right)\label{DeterminantFormula}
\end{equation}
holds for arbitrary square matrices $A$, $B$, $C$, and $D$, if $[A,C]=0$ is satisfied. 
\subsection{Nonmagnetic disorder}\label{sec:standardAndersonTheorem}
We start by considering TRS disorder, i.\,e., we assume $\hat{W}_-=0$ but $\hat{W}_+ \neq 0$ in Eq.~\eqref{DisorderedBdGHam}. From the Anderson theorem we expect the $s^{++}$~state to be robust against such nonmagnetic impurities. The spectrum of $\hat{h}$ can straightforwardly be found from the condition 
\begin{equation}
 \det\begin{pmatrix} \hat{\xi} + \hat{W}^+ -\epsilon\hat{\mathds{1}} & \frac{1}{\sqrt{2}}\left(\Delta_+\hat{\tau}_0 + \Delta_-\hat{\tau}_3\right) \\ \frac{1}{\sqrt{2}}\left(\Delta_+\hat{\tau}_0 + \Delta_-\hat{\tau}_3\right) & -\left(\hat{\xi} + \hat{W}^+ + \epsilon\hat{\mathds{1}}\right) \end{pmatrix} = 0 \mathcomma \label{SCPPAndersonCalc}
\end{equation} 
where for a pure $s^{++}$~pairing state, $\Delta_-=0$ and $\Delta_+\neq0$ holds in addition. 
Then the commutator 
\begin{equation}
 \left[\hat{\xi} + \hat{W}^+ -\epsilon\hat{\mathds{1}},\frac{1}{\sqrt{2}}\left(\Delta_+\hat{\tau}_0 + \Delta_-\hat{\tau}_3\right)\right] = \frac{1}{\sqrt{2}}\Delta_-\left[\hat{W}^+,\hat{\tau}_3\right]  \label{Commutator}
\end{equation} 
vanishes, and we can use Eq.~\eqref{DeterminantFormula} for the evaluation of the determinant. 
We obtain the eigenvalues of $\hat{h}$ in the case of TRS disorder in an $s^{++}$~superconductor, 
\begin{equation}
 \pm \sqrt{(\xi_i+W_i)^2+\Delta_+^2/2} \mathcomma
\end{equation} 
where $\xi_i+W_i$ denote the different real eigenvalues of the Hermitian matrix $\hat{\xi} + W^+$. 
Consequently, the gap of the disordered system is larger than or equal to $|\Delta_+|/\sqrt{2}$, the gap of the clean system. We have hereby shown that the gap will be unaffected by the presence of the disorder potential which indicates the stability of the $s^{++}$~superconducting state against TRS impurity scattering, and thus obtained the Anderson theorem for $s^{++}$~superconductors. 

We note that the commutator~\eqref{Commutator} also vanishes for $\Delta_-\neq0$ if the disorder potential is purely band diagonal, i.\,e., no interband scattering processes occur. Therefore, from similar reasoning, we obtain that the $s^{+-}$~pairing state is protected against nonmagnetic intraband scattering.
\subsection{Anderson theorem for $s^{+-}$~superconductors}\label{sec:generalizedAndersonTheorem}
The same approach can be used to motivate an analog of the Anderson theorem for the $s^{+-}$~pairing  state. We  rewrite the determinant by performing a unimodular transformation in bandspace, 
\begin{equation}
\det\left(\hat{h}-\epsilon\hat{\mathds{1}}\right)=
 \det\left(\begin{pmatrix}\hat{\tau}_0 & 0 \\ 0 &\hat{\tau}_3\end{pmatrix}\left(\hat{h}-\epsilon\hat{\mathds{1}}\right)\begin{pmatrix}\hat{\tau}_0 & 0 \\ 0 & \hat{\tau}_3\end{pmatrix} \right)\mathperiod
\end{equation}
For the specific microscopic scattering mechanism to be discussed below, that is, for a purely band diagonal TRS component and a purely band off-diagonal TRA component of the disorder potential, it follows: 
\begin{equation}
 \hat{\tau}_3 \hat{W}^+ \hat{\tau}_3 = \hat{W}^+, \qquad \hat{\tau}_3 \hat{W}^- \hat{\tau}_3 = -\hat{W}^- \mathcomma
\end{equation} 
and, hence, we have to solve 
\begin{align}
  \det\begin{pmatrix} \hat{\xi} + \hat{W} -\epsilon\hat{\mathds{1}} & \frac{1}{\sqrt{2}}\left(\Delta_+\hat{\tau}_3 + \Delta_-\hat{\tau}_0\right) \\ \frac{1}{\sqrt{2}}\left(\Delta_+\hat{\tau}_3 + \Delta_-\hat{\tau}_0\right) & -\left(\hat{\xi} + \hat{W} + \epsilon\hat{\mathds{1}}\right) \end{pmatrix} =0 \mathperiod\label{DeterminantPreForm}
\end{align}
From the analysis of Sec.~\ref{sec:standardAndersonTheorem} we know that the relevant quantity for the sensitivity to disorder is the commutator
\begin{equation}
 \left[\hat{\xi} + \hat{W} -\epsilon\hat{\mathds{1}},\frac{1}{\sqrt{2}}\left(\Delta_+\hat{\tau}_3 + \Delta_-\hat{\tau}_0\right)\right] = \frac{1}{\sqrt{2}}\Delta_+\left[\hat{W},\hat{\tau}_3\right].
\end{equation} 
It vanishes in case of $s^{+-}$~SC where $\Delta_+ = 0$ and $\Delta_-\neq 0$, but assumes finite values for the $s^{++}$ superconductor when TRA interband scattering is present. This is the algebraic reason for why the $s^{++}$ superconductor is in general prone to TRA scattering, while the $s^{+-}$ state is stable against TRA interband disorder and TRS intraband disorder. 
Let us finally emphasize that this conclusion holds irrespective of the form of the bands (as long as $\xi_{\alpha,\vect{-k}}=\xi_{\alpha,\vect{k}}$ holds) and the detailed momentum dependence of $W_{\alpha,\alpha'}(\vect{k},\vect{k}')$. In particular, the disorder potential does not have to be momentum independent within each band for the $s^{+-}$ Anderson theorem to hold.  
Furthermore, it does not rely  on the disorder potential breaking time-reversal symmetry due to spin or orbital magnetism. It is only important that $\hat{\tau}_3 \hat{W}^\pm \hat{\tau}_3 = \pm \hat{W}^\pm$ holds. Here the insensitivity to spin results from the investigation of singlet pairing.
\subsection{Symmetry protection of density waves}\label{sdw-protection}
Criteria for the stability against a specific class of impurities, analogous to the Anderson theorem for superconductivity, can be derived for particle-hole instabilities as well. The general density wave mean-field Hamiltonian reads 
\begin{align}
 \begin{split}
 \hat{\mathcal{H}}^{\text{MF}}_{\text{DW}} &= \sum_{\vect{k}}\hat{\psi}_{\alpha,\vect{k}}^\dagger \xi_{\alpha,\vect{k}} \hat{\psi}_{\alpha,\vect{k}}^{} + \sum_{\vect{k}}\hat{\psi}_{\alpha,\vect{k}}^\dagger \hat{O}_{\alpha,\alpha'} \hat{\psi}_{\alpha^\prime,\vect{k}}^{}\mathcomma  \\
 \hat{O} &= \begin{pmatrix} 0 & m \\ m^\dagger & 0 \end{pmatrix} 	
 \end{split}
\end{align}
which includes both real ($m^\dagger= m$) and imaginary ($m^\dagger= -m$) spin ($m = \vect{M}\vect{\sigma}$) and charge ($m \propto\sigma_0$) density waves.
For simplicity we assume particle-hole symmetric bands, i.\,e., $\xi_{2,\vect{k}} = -\xi_{1,\vect{k}}$. In that case, the density wave phases are fully gapped already at infinitesimal $m$. However, our arguments can be extended to the case of small deviations from perfect particle-hole symmetry but then we have to assume an order parameter large enough to ensure a fully gapped Fermi surface.  

Again, we consider an arbitrary but fixed disorder realization as given in Eq.~\eqref{eq:disorderhamiltonian1}. The full mean-field Hamiltonian can be written compactly as $\hat{\mathcal{H}}^\mathrm{MF}_\mathrm{DW}+\hat{\mathcal{H}}_\mathrm{dis}=\frac{1}{2}\sum_{\vect{k},\vect{k}^\prime}\sum_{\alpha,\alpha^\prime}\hPsi_\alpha^\dagger(\vect{k})\hat{h}_{\alpha\alpha^\prime}(\vect{k},\vect{k}^\prime)\hPsi_{\alpha^\prime}(\vect{k}^\prime)$ 
with 
\begin{equation}
 \hat{h} = \begin{pmatrix} \hat{\xi}_1 & 0 \\ 0 & -\hat{\xi}_1 \end{pmatrix} + \hat{O} + \hat{W}\mathperiod 
\end{equation} 
The condition under which the gap of the density wave will not be reduced in the presence of impurities is  
\begin{equation}
 \{\hat{W},\hat{O}\} = 0 \mathcomma \label{AndersonDensityWave} 
\end{equation}
and hence the density wave is stable against impurities satisfying the criterion~\eqref{AndersonDensityWave}. The proof is presented in Appendix~\ref{app:anderson-dw}.

In the remainder we apply this result to our model for the iron-based superconductors, where real spin density wave order (SDW) is competing with superconductivity, i.\,e., we consider $m=\vect{M}\hat{\vect{\sigma}}=m^\dagger$. 
In general, this phase is stable against disorder configurations of the form 
\begin{equation}
 \hat{W} = \hat{\tau}_0 \hat{A}_1 + \hat{\tau}_1 \hat{A}_2 + \hat{\tau}_2 \hat{C}_1 + \hat{\tau}_3 \hat{C}_2\mathcomma
\end{equation}
where $\hat{C}$ and $\hat{A}$ denote matrices in spin and momentum space that are commuting and anticommuting with the order parameter $\hat{O}$, respectively. This is satisfied by the choice 
\begin{align}
 \left(A_j\right)_{\vect{k},\vect{k}'} &= \left(\vect{a}^j\right)_{\vect{k},\vect{k}'} \hat{\vect{\sigma}} \mathcomma  \\
 \left(C_j\right)_{\vect{k},\vect{k}'} &=  \left(c^j_0\right)_{\vect{k},\vect{k}'}\hat{\sigma}_0 + \left(\vect{c}^j\right)_{\vect{k},\vect{k}'} \hat{\vect{\sigma}} \mathcomma 
 \end{align}
 where the vectors $\vect{a}^j$ and $\vect{c}^j$ are oriented perpendicular and parallel to the magnetic order parameter, respectively, i.\,e., it holds that $\left(\vect{a}^j\right)_{\vect{k},\vect{k}'} \perp \vect{M}$ and $\left(\vect{c}^j\right)_{\vect{k},\vect{k}'}  \parallel \vect{M}$. 
If we restrict ourselves to spin-independent impurities, this reduces to
\begin{equation}
 \hat{W} = \begin{pmatrix} c_0^2 & i c_0^1 \\ -i c_0^1 & -c_0^2 \end{pmatrix} \mathperiod
\end{equation} 
Therefore, also scenarios where SDW is stable against impurities are conceivable, and in particular, we find that spin density waves are protected against impurities breaking time-reversal symmetry by nucleation of imaginary charge density order which are discussed in Sec.~\ref{example} as an example relevant for iron-based superconductors. However, SDW order is prone to intraband scattering breaking the particle-hole symmetry of the bands ($\hat{W}\propto \hat{\tau}_0$).
\section{Disorder averaging}\label{sec:averaging}
In the following sections, we do not consider spin-magnetic impurities. 
To evaluate physical observables, we use the disorder-averaging
diagrammatic technique\cite{Abrikosov-Gorkov-Dzyaloshinski:MethodsOfQuantumFieldTheory}.
A basic element of this technique is the impurity line 
\begin{align}
  \label{impline}
  \vcenter{\hbox{\includegraphics[height=4em]{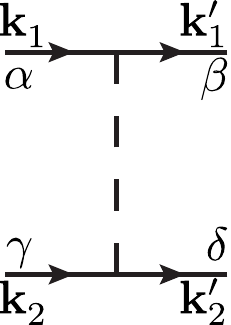}}} 
  &=\sum_{i=1}^N
  \left<
  \left<\vect{k}_1^{},\alpha\right|\hU_{\vect{R}_i}\left|\vect{k}_1^\prime,\beta\right>\left<\vect{k}_2^{},\gamma\right|\hU_{\vect{R}_i}\left|\vect{k}_2^\prime,\delta\right>\right>_{\vect{R}_i} \nonumber\\
  &=(2\pi)^d \Gamma_{\alpha\beta\gamma\delta}(\vect{k}_1^{},\vect{k}_1^\prime,\vect{k}_2^{},\vect{k}_2^\prime) \nonumber \\ 
  & \qquad \times 
  \operatorname{\updelta}(\vect{k}_1^{}+\vect{k}_2^{}-\vect{k}_1^\prime-\vect{k}_2^\prime +\vect{K})\mathcomma
 \end{align}
 where
 \begin{align}
  \Gamma_{\alpha\beta\gamma\delta}(\vect{k}_1^{},\vect{k}_1^\prime,\vect{k}_2^{},\vect{k}_2^\prime)
  &= n_\mathrm{imp}U_{\vect{k}_1^{}\vect{k}_1^\prime}^{\alpha\beta} U_{\vect{k}_2^{}\vect{k}_2^\prime}^{\gamma\delta} \label{eq:scatteringrate} \mathperiod
\end{align}
Here $U_{\vect{k}\vect{k}^\prime}^{\alpha\beta}$ is the matrix element of the perturbation due to a single impurity at site $\vect{R}=\vect{0}$, $\langle\ldots\rangle_{\bR_i}=\Omega^{-1}\int d\bR_i\ldots$
is the averaging with respect to the position $\bR_i$ of impurity $i$, $n_\mathrm{imp}=N/\Omega$ denotes the impurity concentration, and $\Omega$ is the $d$-dimensional volume. It holds that $\bK=0$ if all or two of the momenta $\bk_1$, $\bk_2$, $\bk_1^\prime$, $\bk_2^\prime$
belong to the same band, and $\bK=\bQ$ if one momentum belongs to one band, and three other to the other band. [In Eq.~(\ref{impline}) we have taken into account that $2\bQ$ is a reciprocal vector].

\begin{figure}[b]
  \includegraphics[width=0.8\columnwidth]{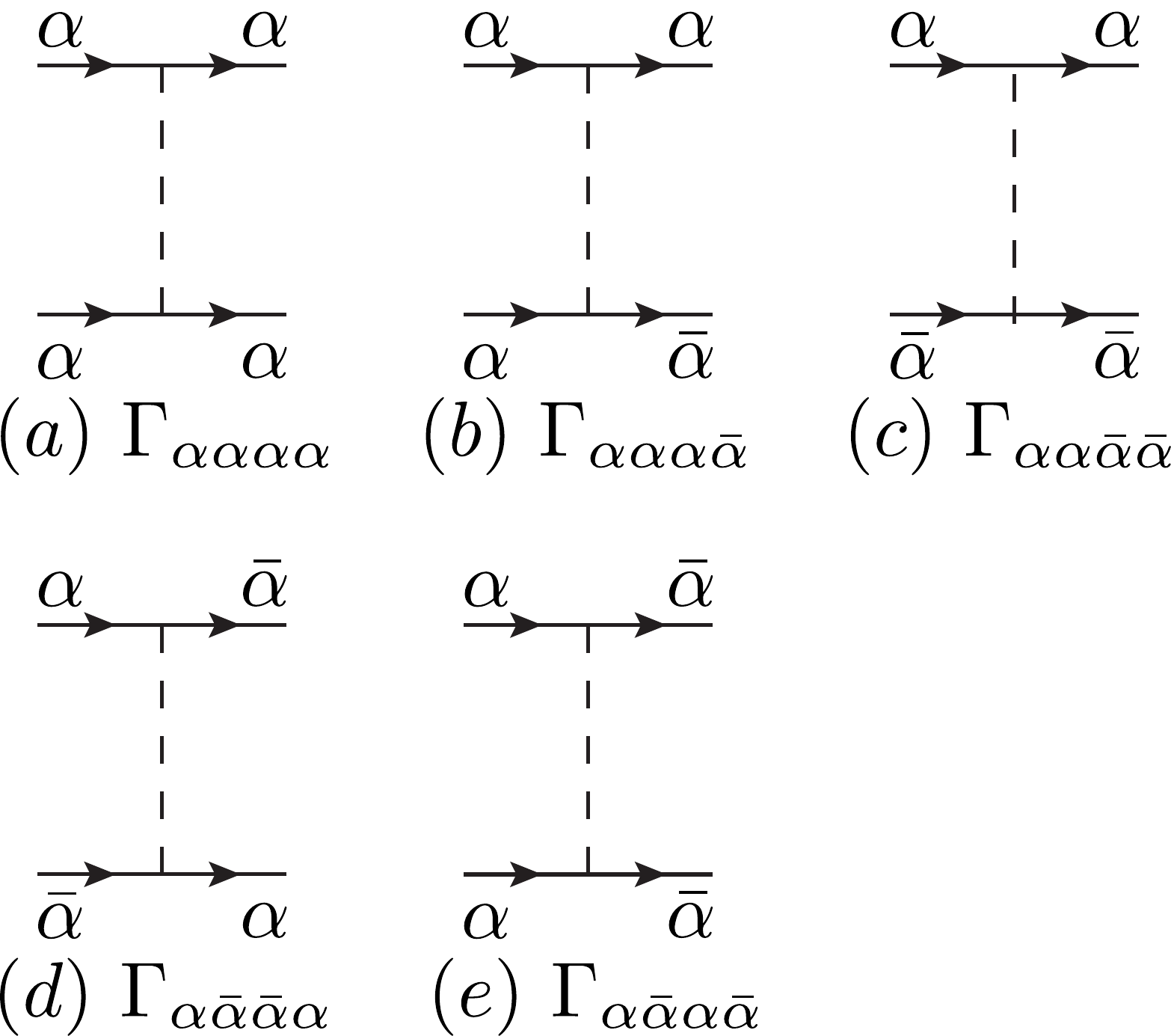}
  \caption{Scattering processes that can occur in a two-band system. (a) Intraband scattering process in band~$\alpha$. (b)--(e) Interband scattering processes. }
  \label{fig:scattering_times}
\end{figure}

The impurity line, Eq.~\eqref{impline}, describes the elastic scattering of two momentum states $\bk_1$ and
$\bk_2$ into another two momentum states $\bk_1^\prime$ and $\bk_2^\prime$. The scattering can occur 
within the same band or involve interband processes, as shown in Fig.~\ref{fig:scattering_times}.
The $\updelta$~function in Eq.~\eqref{impline} represents the conservation of quasimomentum, and
the quantity $\Gamma_{\alpha\beta\gamma\delta}(\vect{k}_1^{},\vect{k}_1^\prime,\vect{k}_2^{},\vect{k}_2^\prime)$, defined in Eq.~\eqref{eq:scatteringrate}, is hereinafter referred to as the rate of elastic scattering between the pair of momentum states $\bk_1$, $\bk_2$
and $\bk_1^\prime$, $\bk_2^\prime$, respectively.

The {\it intraband scattering} process within band~$\alpha$ is depicted in Fig.~\ref{fig:scattering_times}(a), and we abbreviate the corresponding scattering rate by $\Gamma_\alpha\equiv\Gamma_{\alpha\alpha\alpha\alpha}$. For sufficiently short-correlated disorder considered in this paper, the rates 
\begin{equation}
   \Gamma_{1}\approx n_\mathrm{imp}|U_{\bnil\bnil}^{11}|^2 \mathcomma \quad 
   \Gamma_{2}\approx n_\mathrm{imp}|U_{\vect{Q}\vect{Q}}^{22}|^2
\end{equation}
are independent of the momenta $\bk_1$, $\bk_2$, $\bk_1^\prime$, and $\bk_2^\prime$. 
Such intraband scattering processes are  pair-breaking neither for conventional nor for 
unconventional superconducting states.
We emphasize that in general $\Gamma_1\neq\Gamma_2$, because the momentum states
in the two bands may have different structure, e.g., in terms of sublattices or atomic orbital
degrees of freedom, and thus may be scattered differently by impurities.

Processes involving {\it interband scattering} are shown in Figs.~\ref{fig:scattering_times}(b)--(e). 
The process in Fig.~\ref{fig:scattering_times}(b) requires a momentum transfer of $\vect{K}=\vect{Q}$ which is not a reciprocal lattice vector and thus this scattering process is forbidden due to the conservation of quasimomentum. 
The process in Fig.~\ref{fig:scattering_times}(c) affects neither the quasiparticle self-energy part nor the superconductive properties but can be important, e.g., for the magnetic properties of the material.
The process depicted in Fig.~\ref{fig:scattering_times}(d) affects the quasiparticle self-energy part, as we discuss in Sec.~\ref{Sec:selfenergy}.
In what follows, we assume that the respective rate $\Gamma_{\alpha\bar{\alpha}\bar{\alpha}\alpha}$
is independent of the momenta $\bk_1$, $\bk_2$, $\bk_1^\prime$, and $\bk_2^\prime$
and, generally speaking, is different from $\Gamma_1$ and $\Gamma_2$.
Such assumption is rather generic and may be justified, e.g., 
if the disorder (perturbation $\hU_\bR$) has components varying both on length scales significantly
smaller than $1/|\bQ|$ and on scales $\lambda$: $1/|\vect{Q}|\ll \lambda\ll1/k_\mathrm{F}$.
The former will contribute to the intraband scattering rates as well as to the interband
scattering rates, whereas the latter contributes significantly only to the interband scattering rates.
The rate of the process in Fig.~\ref{fig:scattering_times}(d) is
\begin{equation}
 \label{G1221}
 \Gamma_{\alpha\bar{\alpha}\bar{\alpha}\alpha}
 \approx n_\mathrm{imp}|U_{\vect{0}\vect{Q}}^{\alpha\bar{\alpha}}|^2.
\end{equation}
We note that the rate given in \eqref{G1221} is real, $\Gamma_{\alpha\bar{\alpha}\bar{\alpha}\alpha}\in\mathds{R}$, and it holds that $\Gamma_{1221}=\Gamma_{2112}$.
On the contrary, the scattering process shown in Fig.~\ref{fig:scattering_times}(e) in general comes with a phase
\begin{equation}
 \Gamma_{\alpha\bar{\alpha}\alpha\bar{\alpha}}
 \approx n_\mathrm{imp}(U_{\vect{0}\vect{Q}}^{\alpha\bar{\alpha}})^2=n_\mathrm{imp}|U_{\vect{0}\vect{Q}}^{\alpha\bar{\alpha}}|^2 \mathrm{e}^{\mathrm{i}\phi_\alpha}\mathcomma
 \label{GammaPhase}
\end{equation}
where $\phi_\alpha\neq0$ (modulo $2\pi$) if $\Im U_{\vect{0}\vect{Q}}^{\alpha\bar{\alpha}}\neq0$. 
This process describes the scattering of a pair of momentum states in one band into
a pair of momentum states in the other band. 
Since $|\Gamma_{\alpha\bar{\alpha}\alpha\bar{\alpha}}|=\Gamma_{\alpha\bar{\alpha}\bar{\alpha}\alpha}$,
we introduce the notation $\Gamma_{12}\equiv\Gamma_{1221}\in\mathds{R}$ and the phase~$\phi$, 
\begin{equation}
  \Gamma_{1212}=\Gamma_{12} \mathrm{e}^{\mathrm{i}\phi} \mathcomma \quad 
  \Gamma_{2121}=\Gamma_{12} \mathrm{e}^{-\mathrm{i}\phi} \mathperiod
\end{equation}

In principle, the phase $\phi$
is defined relative to a similar phase of the BCS coupling matrix element~$V^{\alpha\bar{\alpha}}_{\vect{k},\vect{k}^\prime}$, that is contained in Eq.~\eqref{eq:hamiltonian-int}, which couples pairs of momentum states in different bands.
Thus, the interplay of the scattering process in Fig.~\ref{fig:scattering_times}(e) and the superconductive
coupling may affect the superconductive properties of the system. 
The fact that $\phi$ must be understood as a relative phase becomes more evident in our discussion in Sec.~\ref{sec:conclusion}. 
\subsection*{Self energy and Cooperons} \label{Sec:selfenergy}
Assuming that the scattering is sufficiently weak such that the mean-free path $l=v_\mathrm{F}\tau$ 
satisfies $k_\mathrm{F}l\gg 1$, single-particle interference effects are subleading. This means, diagrams with crossed impurity lines can be neglected since they are suppressed by a factor $1/k_\mathrm{F}l$.  

Because the process in Fig.~\ref{fig:scattering_times}(b) is forbidden by quasimomentum conservation,
only the processes in Figs.~\ref{fig:scattering_times}(a) and~\ref{fig:scattering_times}(d) contribute
to the electron self-energy part in the Born approximation, 
\begin{equation}
  \Sigma_\alpha=\vcenter{\hbox{\includegraphics[height=3em]{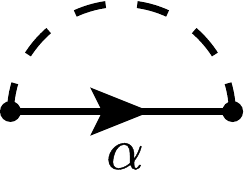}}}+\vcenter{\hbox{\includegraphics[height=3em]{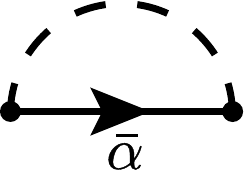}}} \mathcomma 
\end{equation} 
and therefore, in the disorder-averaged electron propagator, 
\begin{align}
\begin{split}
  \label{Propagator}
 G_{\alpha,\vect{k}}(\nu_n)&=
 \frac{1}{\mathrm{i}\nu_n-\xi_{\alpha,\vect{k}}+\frac{\mathrm{i}}{2\tau_{\alpha}}\sgn\nu_n}\mathcomma \\
 \tau_\alpha&=\left[2\pi\rho_\mathrm{F}\left(\Gamma_\alpha+\Gamma_{12}\right)\right]^{-1} \mathcomma 
 \end{split}
\end{align}
the full scattering rate that determines the elastic scattering time~$\tau_{\alpha}$ in band~$\alpha$ is a sum of the intraband~$\Gamma_{\alpha}$ and
the interband~$\Gamma_{12}$ rates. 

\begin{figure}
  \includegraphics[width=0.8\columnwidth]{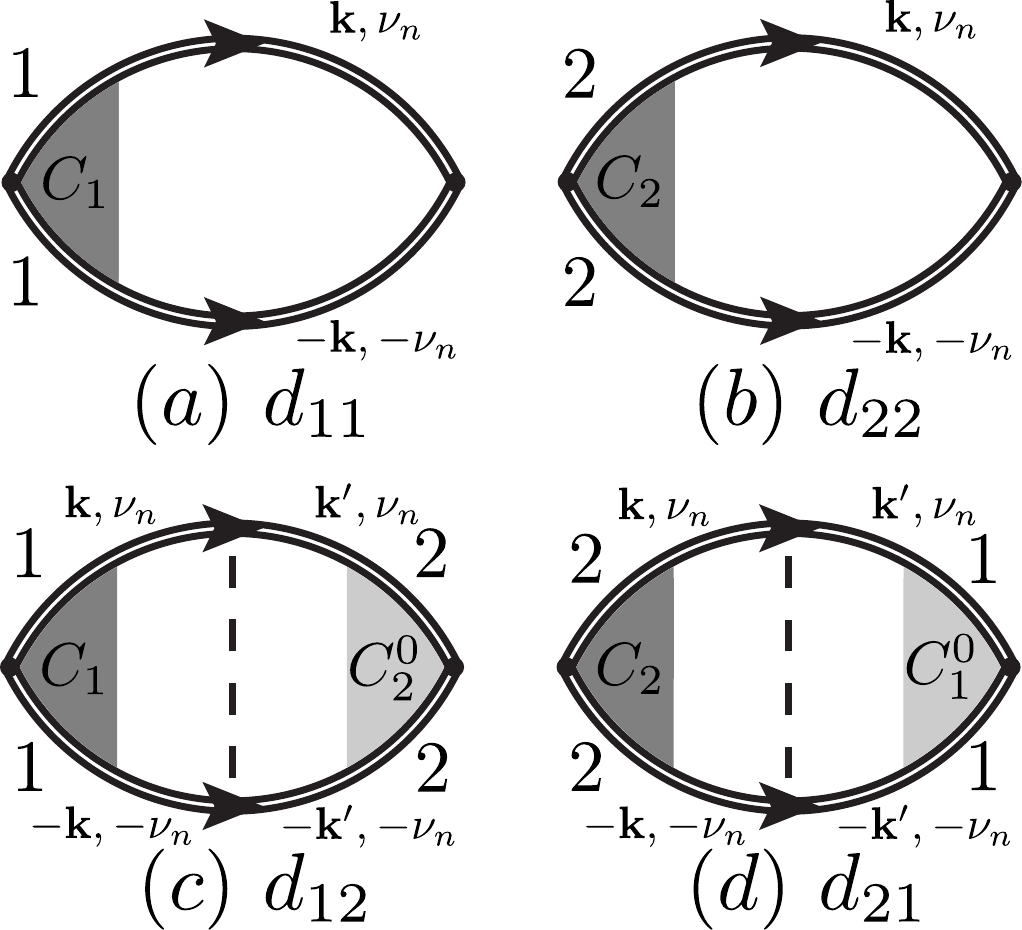}
  \caption{Diagrams that contribute to the quadratic coefficients of the free energy in the presence of intraband and interband scattering. Here dark gray vertices indicate vertex renormalization given by the generalized Cooperon ladder, whereas light gray vertices are only renormalized by the respective single-band Cooperon ladder. Diagrams (a) and (b) also survive in the absence of interband scattering, whereas diagrams (c) and (d) are only nonzero if interband scattering processes are included.}
  \label{fig:diagrams}
\end{figure}

Further corrections due to impurity scattering can be conveniently summarized into vertex corrections, as they appear in the diagrams contributing to the SC~transition temperature shown in Fig.~\ref{fig:diagrams}.
In the presence of intraband as well as interband scattering processes, most contributions can be accounted for by a generalized form~$C_{\alpha}$ of the Cooperon ladder of the impurity line. This generalized Cooperon is indicated in dark gray in the diagrams in Fig.~\ref{fig:diagrams} and accounts for all combinations of scattering processes starting and ending in band~$\alpha$, including (pairwise) interband scattering processes and intermediate scattering processes in band~$\bar{\alpha}$. 
To calculate this generalized Cooperon, a single rung of the Cooperon ladder in band~$\alpha$ as known from one-band models has  to be modified as 
\begin{align}
  \vcenter{\hbox{\includegraphics[height=4em]{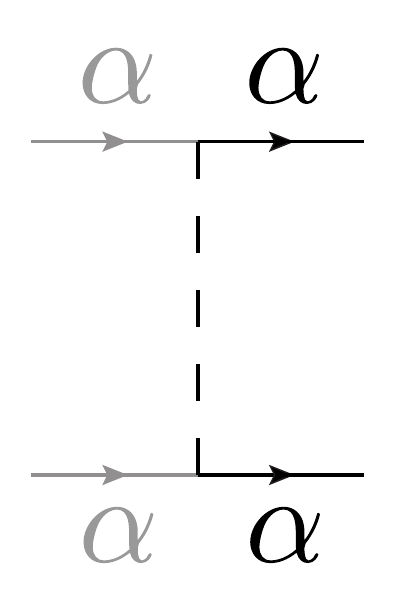}}} & \rightarrow\vcenter{\hbox{\includegraphics[height=4em]{./cooperon1_rev}}}    +\vcenter{\hbox{\includegraphics[height=4em]{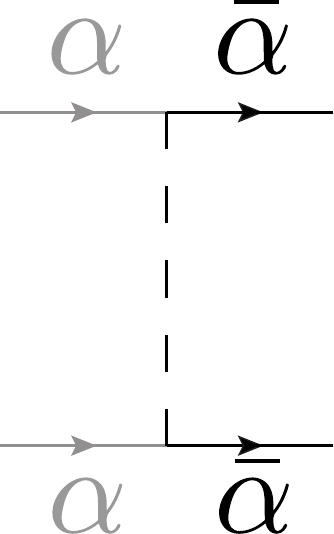}}} \times\Bigg(1+\vcenter{\hbox{\includegraphics[height=4em]{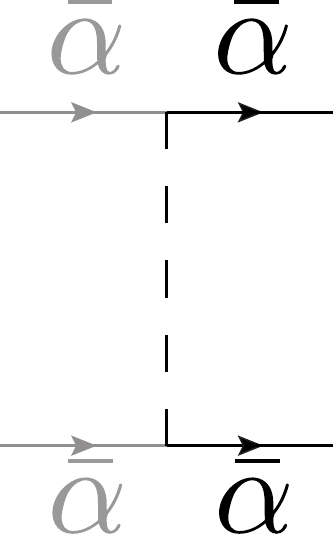}}} \nonumber \\ 
  &\qquad \qquad+\vcenter{\hbox{\includegraphics[height=4em]{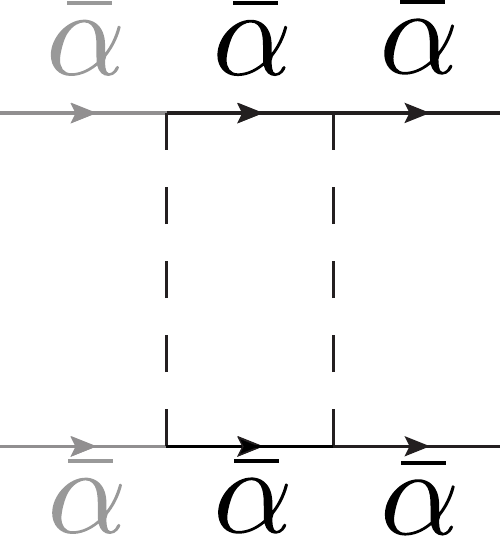}}}+\ldots\Bigg)\times\vcenter{\hbox{\includegraphics[height=4em]{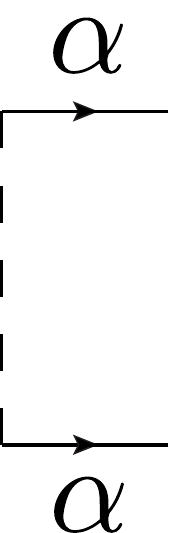}}} \mathcomma
\end{align}
where the gray lines do not enter the calculation but are drawn for the sake of clarification. 
The summation of the full ladder is presented in Appendix~\ref{app:cooperon-ladder} and leads to a frequency-dependent factor 
\begin{equation}
  C_\alpha(\nu_n)=\frac{(\pi\rho_\mathrm{F}\Gamma_{12}+|\nu_n|)(\pi\rho_\mathrm{F}(\Gamma_{\alpha}+\Gamma_{12})+|\nu_n|)}{|\nu_n|(2\pi\rho_\mathrm{F}\Gamma_{12}+|\nu_n|)}
\end{equation}
at vertices associated with the order parameter $\Delta_\alpha$.  
We note that the vertex corrections $C_\alpha(\nu_n)$ associated with $\Delta_\alpha$ only depend on the intraband scattering rate in the respective band~$\alpha$, and on the interband scattering rate $\Gamma_{12}$. The vertex corrections are independent of the other band to which electrons are scattered in intermediate processes and within which they can also be scattered. 

In addition, to avoid double counting in the interband diagrams $d_{12}$ and $d_{21}$, we need the usual single-band Cooperon ladder in band~$\alpha$, $C_{\alpha}^0$, which is indicated in light gray in the diagrams in Fig.~\ref{fig:diagrams}, and given by 
\begin{equation}
  C^0_\alpha(\nu_n)=\frac{|\nu_n|+\pi\rho_\mathrm{F}(\Gamma_\alpha+\Gamma_{12})}{|\nu_n|+\pi\rho_\mathrm{F}\Gamma_{12}}\mathperiod \label{eq:usualCooperon}
\end{equation}
Note that both intraband $\Gamma_\alpha$ and interband $\Gamma_{12}$ scattering rates enter the Cooperon in Eq.~\eqref{eq:usualCooperon} through the disorder-averaged electron propagators as given in Eq.~\eqref{Propagator}.

Furthermore, even though it is the elastic scattering process
in Fig.~\ref{fig:scattering_times}(e), associated with a nontrivial phase factor, see Eq.~\eqref{GammaPhase}, that is accounted for by the above vertex correction, this process enters only pairwise with its complex conjugate, and thus the resulting vertex corrections are real. 
Therefore, all physical observables which contain only electron self energies~$\Sigma_\alpha$ and vertex corrections~$C_{\alpha}$ and~$C_{\alpha}^0$ 
are unaffected by the phase factor arising in the interband scattering process that is defined in Fig.~\ref{fig:scattering_times}(e) and Eq.~\eqref{GammaPhase}. 
However, not all contributions arising from impurity scattering can be summarized in terms of electron self energies and vertex corrections, and consequently, physical observables can indeed be affected by such a phase related to orbital magnetism, the most prominent example for superconductors being the superconducting transition temperature~$T_\mathrm{c}$, as established in Sec.~\ref{sec:tc}.
\section{Transition temperature in the presence of impurity scattering}\label{sec:tc}
The action associated with the interacting Hamiltonian given in Eq.~\eqref{eq:hamiltonian-int} can be decoupled by introduction of the auxiliary fields $\Delta_\pm=\frac{1}{\sqrt{2}}(\Delta_1\pm\Delta_2)$, where $\Delta_1$ and $\Delta_2$ are the values of the order parameter on the respective sheets of the Fermi surface. For attractive interaction we use 
\begin{align}
\begin{split}
  \mathrm{e}^{-Vb_+^\ast b_+^{}}&=\int\mathcal{D}\Delta_+^\ast\mathcal{D}\Delta_+^{}\,\mathrm{e}^{\frac{1}{V}\Delta_+^\ast \Delta_+^{}+\Delta_+^\ast b_+^{}+\Delta_+^{}b_+^\ast} \mathcomma \\
  \mathrm{e}^{Vb_-^\ast b_-^{}}&=\int\mathcal{D}\Delta_-^\ast\mathcal{D}\Delta_-^{}\,\mathrm{e}^{\frac{1}{V}\Delta_-^\ast\Delta_-^{}+\mathrm{i}\Delta_-^\ast b_-^{}+\mathrm{i}\Delta_-^{}b_-^\ast}\mathcomma
\end{split}
\end{align}
where $b_\pm=\tfrac{1}{\sqrt{2}}(b_1\pm b_2)$ which are linked to the original fermionic fields by $b_\alpha=\sum_{\vect{k}}\psi_{\alpha,-\vect{k},\downarrow}\psi_{\alpha,\vect{k},\uparrow}$. The respective decoupling for repulsive interaction has the same structure, but the factor $\mathrm{i}$ is then associated with the $\Delta_+$~mode rather than the $\Delta_-$~mode, ensuring the convergence of the integral. 

Then the SC~transition temperature can be extracted from the quadratic part of an expansion of the free energy in terms of the order parameters $\Delta_+$ and $\Delta_-$, which can be written in matrix form as   
\begin{equation}
  \Delta F=\left(\begin{matrix}\Delta_+^\ast & \Delta_-^\ast\end{matrix}\right)\left(\begin{matrix}a_{++} & a_{+-} \\ a_{-+} & a_{--}\end{matrix}\right)\left(\begin{matrix} \Delta_+ \\ \Delta_-\end{matrix}\right)\mathperiod 
  \label{eq:expansion-quadratic-part}
\end{equation}
The sign change of the lower eigenvalue of this quadratic form, 
\begin{align}
  \lambda_{1,2}&=\frac{1}{2}(a_{++}+a_{--}) \nonumber \\ 
  &\qquad \pm\frac{1}{2}\sqrt{(a_{++}-a_{--})^2+4a_{+-}a_{-+}} \mathcomma 
\end{align}
determines the transition temperature. 
The coefficients in this expansion of the free energy in the presence of disorder can be obtained from our microscopic model, and the intraband and interband diagrams $d_{ij}$ contributing to the quadratic coefficients are depicted in Fig.~\ref{fig:diagrams}. 

\begin{figure}[t]
  \includegraphics[width=0.9\columnwidth]{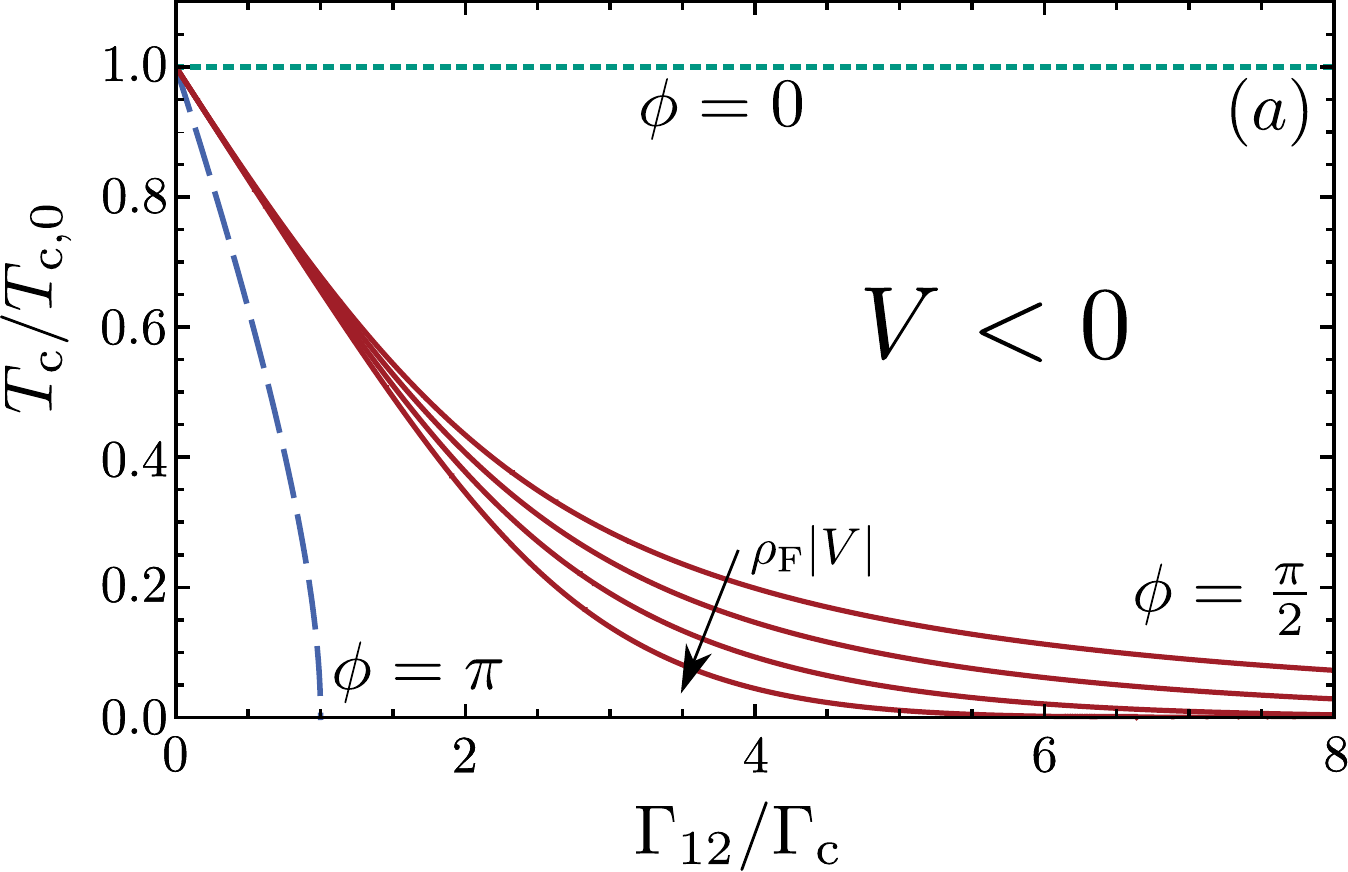}\\[2em]
  \includegraphics[width=0.9\columnwidth]{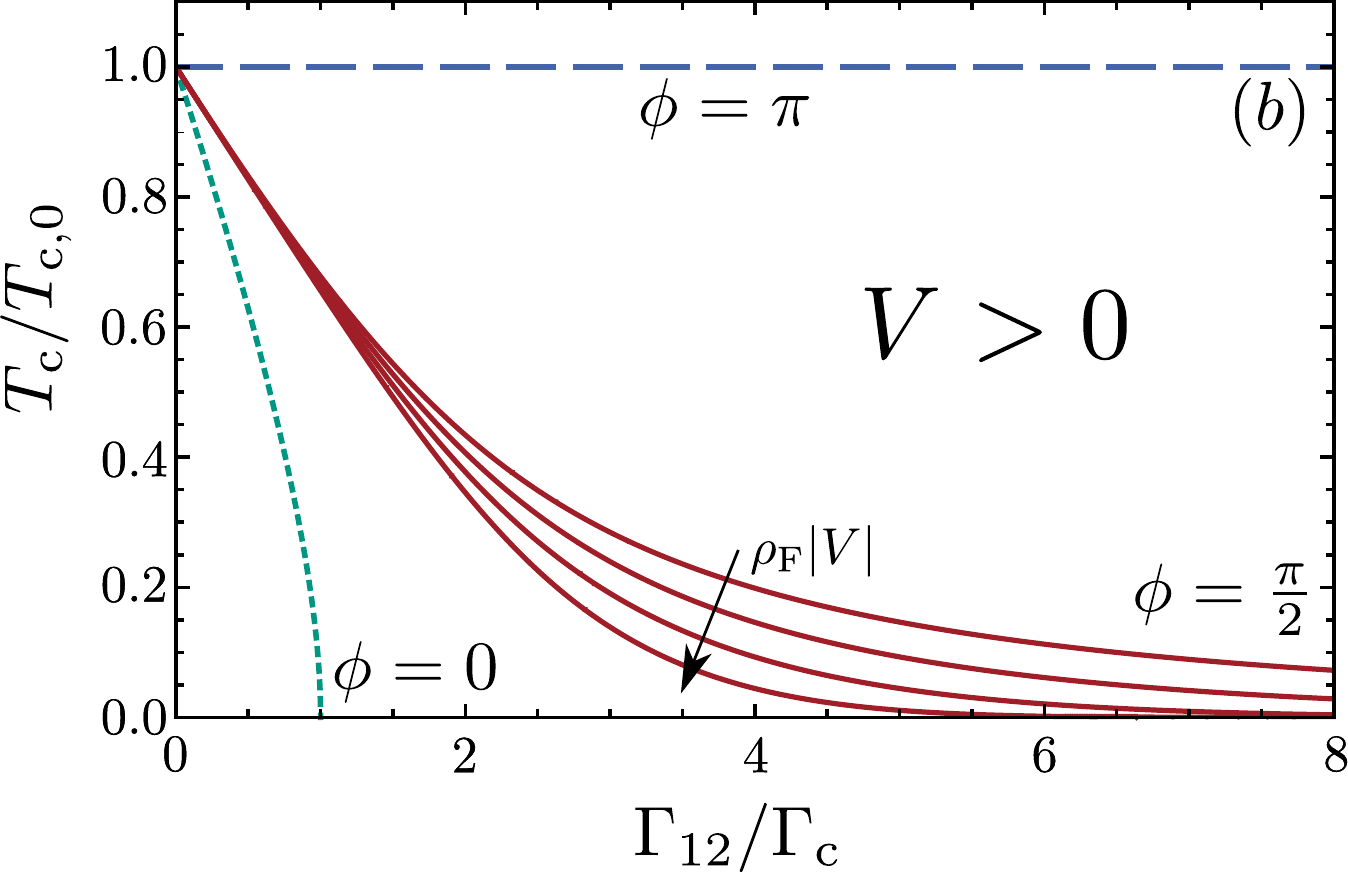}
  \caption{(Color online) Suppression of the transition temperature $T_\mathrm{c}$ with increasing interband scattering rate ${\Gamma_{12}}$ for phases~$\phi=0$ (green dotted line), $\phi=\frac{\pi}{2}$ (red lines), and $\phi=\pi$ (blue dashed line) in case of (a)~attractive and (b)~repulsive interaction. For $\phi=\frac{\pi}{2}$, the transition temperature depends on the dimensionless coupling constant, and we plotted our results for $\rho_\mathrm{F}|V|\in\{0.1,0.2,0.3,0.4\}$.}
  \label{fig:suppression}
\end{figure}

The quadratic coefficients in terms of these diagrams read
\begin{align}
  a_{++}&=\frac{1}{|V|}+\frac{1}{2}\sgn V\left[d_{11}+d_{22}+d_{12}+d_{21}\right] \mathcomma\\
  a_{--}&=\frac{1}{|V|}-\frac{1}{2}\sgn V\left[d_{11}+d_{22}-d_{12}-d_{21}\right] \mathcomma\\ 
  a_{+-}&= -\frac{\mathrm{i}}{2}\left[d_{11}-d_{22}+d_{12}-d_{21}\right] \mathcomma\\ 
  a_{-+}&=-\frac{\mathrm{i}}{2}\left[d_{11}-d_{22}-d_{12}+d_{21}\right]\mathperiod
\end{align}
Since for equal density of states in the two bands, $d_{11}=d_{22}$ and $d_{12}=d_{21}^\ast$, the eigenvalues reduce to 
\begin{equation}
  \lambda_{1,2}=\frac{1}{|V|}+\sgn V\Re d_{12}\pm \sqrt{d_{11}^2-(\Im d_{12})^2}\mathcomma 
\end{equation}
and the sign change of the lower one determines the transition temperature. The respective diagrams can be evaluated analytically, and expressed in terms of digamma functions~$\psi_0$, 
\begin{align}
  d_{11}&=d_{22} = \frac{\dosf}{2}\left[\operatorname{\psi_0}\left(\tfrac{1}{2}+\tfrac{\Lambda}{2\pi T}\right)-\operatorname{\psi_0}\left(\tfrac{1}{2}\right)\right. \nonumber \\ 
  & \qquad +\left.\operatorname{\psi_0}\big(\tfrac{1}{2}+\tfrac{\Lambda}{2\pi T}+\tfrac{\rho_\mathrm{F}\Gamma_{12}}{T}\big)-\operatorname{\psi_0}\big(\tfrac{1}{2}+\tfrac{\rho_\mathrm{F}\Gamma_{12}}{T}\big)\right] \nonumber \\
  &= \frac{\dosf}{2}\left\{\begin{matrix} 2\ln\frac{\Lambda}{2\pi T}, & {T}\gg {\rho_\mathrm{F}\Gamma_{12}}\mathcomma \\ \ln\big(\frac{\Lambda^2}{(2\pi)^2\rho_\mathrm{F}\Gamma_{12} T}\big)-\operatorname{\psi_0}\left(\frac{1}{2}\right), & {T}\ll {\rho_\mathrm{F}\Gamma_{12}}\mathcomma \end{matrix} \right. \\
  d_{12}&=d_{21}^\ast=
 \frac{\dosf}{2}\mathrm{e}^{\mathrm{i}\phi}\left[\operatorname{\psi_0}\big(\tfrac{1}{2}+\tfrac{\rho_\mathrm{F}\Gamma_{12}}{T}\big)-\operatorname{\psi_0}\left(\tfrac{1}{2}\right)\right] \nonumber \\
 &=\frac{\rho_\mathrm{F}}{2}\mathrm{e}^{\mathrm{i}\phi}\left\{\begin{matrix}\frac{\pi^2}{2}\frac{\rho_\mathrm{F}\Gamma_{12}}{T}, & T\gg \rho_\mathrm{F}\Gamma_{12}\mathcomma \\ \ln\big(\frac{\rho_\mathrm{F}\Gamma_{12}}{T}\big)-\operatorname{\psi_0}\left(\frac{1}{2}\right), & T\ll \rho_\mathrm{F}\Gamma_{12}\mathcomma \end{matrix}\right. 
\end{align}
where we also gave the results in the limiting cases of a clean system and strong interband scattering (but in the sense that $1/k_\mathrm{F}l\ll 1$ still holds). 
The transition temperature can be determined numerically from these diagrams for arbitrary phases of $\phi$, but it is most instructive to highlight three important limits, namely $\phi=0$, $\phi=\frac{\pi}{2}$, and $\phi=\pi$. 
Our results for the SC~transition temperature as a function of the interband scattering rate are shown in Fig.~\ref{fig:suppression} for attractive and repulsive interaction.

In the clean case and for $\phi=0$, we reproduce well-known results, namely that, depending on the sign of the coupling constant $V$, one of the two modes condenses. 
In case of attractive interaction, $s^{++}$~superconductivity, characterized by the order parameter $\Delta_+$, is realized, whereas for repulsive interaction, it is $s^{+-}$ superconductivity characterized by $\Delta_-$. 
The SC~transition occurs at the critical temperature $T_{\mathrm{c},0}$, as known from BCS~theory, 
\begin{equation}
  T_{\mathrm{c},0}=\Lambda\frac{2\mathrm{e}^\gamma}{\pi}\mathrm{e}^{-\frac{1}{|V|\rho_\mathrm{F}}}\mathcomma
\end{equation}
where $\gamma$ denotes the Euler constant. 
Furthermore, the consideration of $\phi=0$ in a dirty superconductor is also consistent with previous work. 
In case of attractive interaction, the $\Delta_+$~mode condenses, and the transition temperature is unaffected by the presence of impurities, $T_{\mathrm{c}}\approx T_{\mathrm{c},0}$. 
This result for $s^{++}$~SC is known as the Anderson theorem and, as expected, consistent with our symmetry analysis of Sec.~\ref{sec:standardAndersonTheorem}. 
For repulsive interaction we find the $\Delta_-$~mode to be the one that condenses, and now (unconventional) SC is affected by the presence of impurities, and the suppression of the transition temperature is given by the usual Abrikosov-Gorkov law\cite{Soviet.Phys.JETP.12.1243}. Particularly, at a critical scattering rate 
\begin{equation}
 \Gamma_{\mathrm{c}}=\frac{T_{\mathrm{c},0}}{4\mathrm{e}^\gamma\rho_\mathrm{F}}\mathcomma
\end{equation}
$s^{+-}$~superconductivity vanishes completely. 

However, for a phase of $\pi$, we find the reversed situation: Conventional superconductivity is now harmed by  impurities, and even suppressed at a critical scattering rate, whereas for $s^{+-}$~SC, there exists an analog of the Anderson theorem as also follows from our symmetry analysis in Sec.~\ref{sec:generalizedAndersonTheorem}. An illustration of these results can be found in Figs.~\ref{fig:visualization}(b) and ~\ref{fig:visualization}(c).

In the case of $\phi=\frac{\pi}{2}$, we find that the transition temperature is suppressed for attractive as well as repulsive interaction. However, in neither case, a critical scattering rate at which superconductivity vanishes is found, 
\begin{equation}
 T_\mathrm{c}=\left\{\begin{matrix} T_{\mathrm{c},0}-\frac{\pi^2}{4}\rho_\mathrm{F}\Gamma_{12}\mathcomma & T\gg \rho_\mathrm{F}\Gamma_{12}\mathcomma \\ \Lambda\frac{2\mathrm{e}^\gamma}{\pi}\mathrm{e}^{-\frac{1}{\rho_\mathrm{F}V_{\mathrm{eff}}(\Gamma_{12})}} \mathcomma & T\ll \rho_\mathrm{F}\Gamma_{12}\mathcomma \end{matrix}\right.
\end{equation}
where $V_{\mathrm{eff}}(\Gamma_{12})=\dosf|V|^2\ln(\frac{\Lambda}{2\pi\rho_\mathrm{F}\Gamma_{12}})$. Furthermore, the pairing state in case of such an intermediate phase is a superposition of the $\Delta_+$ and $\Delta_-$~mode.

Studies considering SDW order coexisting with superconductivity in iron-based superconductors\cite{FernandesVavilovChubukov2012,LiXu2010} found that the superconducting transition temperature can increase with increasing disorder in the underdoped regime. This is due to the fact that SDW order is affected more severely by impurity scattering than superconductivity. Based on our analysis of Sec.~\ref{sdw-protection}, we expect that such a behavior occurs in the case of dominant particle-hole symmetry-breaking intraband scattering. On the other hand, for interband scattering due to imaginary charge density wave impurities as discussed in Sec.~\ref{example}, we find that both SDW as well as $s^{+-}$~superconductivity are protected. 
\section{Application to iron-based superconductors}
We showed that $s^{++}$~superconductivity can be destroyed by impurities which cause certain interband scattering processes characterized by a nontrivial phase in the impurity line, whereas the $s^{+-}$~pairing state remains robust under certain conditions. In this section we establish the connection of our preceding observations to the situation in iron-based superconductors. We reveal the necessity of time-reversal-symmetry breaking for the occurrence of the effect in these materials and discuss the nucleation of imaginary charge density wave (iCDW) order around impurities as a possible origin of time-reversal-symmetry breaking associated with orbital magnetism in these materials. 
\subsection{Role of time-reversal symmetry}
As anticipated in Sec.~\ref{sec:generalizedAndersonTheorem}, the consideration of time-reversal symmetry-breaking interband scattering allows us to formulate an analog of the Anderson theorem for $s^{+-}$~superconductivity. This was formalized in Sec.~\ref{sec:averaging} by the introduction of a nontrivial phase in the interband scattering rate $\Gamma_{1212}$. 

In this section we elucidate the role of time-reversal symmetry-breaking impurities in iron-based superconductors, where electrons from $d$~orbitals\cite{Graser2009,Zhang2009} are forming the superconducting condensate.

Since $\Gamma_{\alpha\bar{\alpha}\alpha\bar{\alpha}}\propto (U_{\vect{0}\vect{Q}}^{\alpha\bar{\alpha}})^2$, in order to have a nontrivial phase in the impurity line, we need $U_{\vect{0}\vect{Q}}^{\alpha\bar{\alpha}}=\sum_{s,s^\prime}\mathrm{e}^{-\mathrm{i}\vect{R}_{s^\prime}\cdot\vect{Q}}J_{ss^\prime}^{\alpha\bar{\alpha}}$ to have a nonzero imaginary part. Since $2\vect{Q}$ is a reciprocal lattice vector, and thus  $\exp(-\mathrm{i}\vect{R}\cdot\vect{Q})=\pm1$ for any lattice vector $\vect{R}$, this requirement can only be met if the matrix element 
\begin{equation}
  J_{ss^\prime}^{\alpha\bar{\alpha}}
  =\int\mathrm{d}\vect{r}\,(\varphi_{\vect{R}_s}^\alpha(\vect{r}))^\ast U_{\vect{R}=\vect{0}}^{}\varphi_{\vect{R}_{s^\prime}}^{\bar{\alpha}}(\vect{r}) 
\end{equation}
itself has a nonzero imaginary part.
Here, $\varphi_{\vect{R}_s}^\alpha(\vect{r})$ denotes the Wannier function of band~$\alpha$ centered around site~$\vect{R}_s$. The Wannier functions in band space are related to the tight-binding wave functions in orbital space by an orthogonal, that is, real, transformation matrix, since the dispersion in band space is symmetric.

The wavefunctions of electrons on $d$~orbitals with which we are concerned in the iron-based superconductors, can be chosen real, so $J_{ss^\prime}^{\alpha\bar{\alpha}}$ can have an imaginary part only due to the phases in the impurity Hamiltonian. 

In the absence of spin-orbit coupling, the Hamiltonian can
be split into an orbital and a spin part,  $\mathcal{\hat H}_\mathrm{imp}
=\mathcal{\hat H}_\mathrm{imp}^\mathrm{orb}\otimes\mathcal{\hat H}_\mathrm{imp}^\mathrm{spin}$. We consider the transformation properties under time reversal, described by the operator 
\begin{equation}
  \hT=(\hT^\mathrm{orb}\otimes \hT^\mathrm{spin})\mathcal{\hK}\mathcomma
\end{equation}
where $\hat{\mathcal{K}}$ denotes complex conjugation. For spin-$\frac{1}{2}$, the spin part $\hT^\mathrm{spin}$ is given by the Pauli matrix $\mathrm{i}\hat{\sigma}_2$. In real space, the orbital part $\hT^\mathrm{orb}$ is just the identity, $\hT^\mathrm{orb}=\hat{\mathds{1}}^\mathrm{orb}$. 

We consider the most generic time-reversal symmetric impurity Hamiltonian $\hat{\mathcal{H}}_\mathrm{imp}=\hT \hat{\mathcal{H}}_\mathrm{imp} \hT^{-1}$, and if $\hat{\mathcal{H}}_\mathrm{imp}$ is invariant under time reversal, the matrix element $J_{ss^\prime}^{\alpha\bar{\alpha}}$ is invariant as well. If we do not consider scattering processes involving spin flips, that is, if $\hat{\mathcal{H}}_\mathrm{imp}^\mathrm{spin}\propto \hat{\sigma}_0$, then the spin part is also invariant under time reversal, and as a consequence, the orbital part of the Hamiltonian is real, yielding $J_{ss^\prime}^{\alpha\bar{\alpha}}\in\mathds{R}$. 

In conclusion, impurities that are invariant under time-reversal are not able to generate nontrivial phases in the scattering matrix elements such that a nontrivial phase can arise. Since we are not concentrating on spin magnetism, this implies that a nontrivial phase is caused by orbital magnetism in multiband superconductors.
\subsection{iCDW impurities}\label{example}
The renormalization group analysis~\cite{Chubukov2008} of the two-band Hubbard model with particle-hole symmetry, as suited for the description of iron-based superconductors, revealed the existence of a fixed point where the Hamiltonian exhibits an SO(6) symmetry, and three different states of order compete\cite{Chubukov2008,Podolsky2009,KangTesanovic2011}. For repulsive interband interactions, these ordered states are spin density waves (SDW) with a real order parameter
\begin{equation}
 \mathbf{M} = \sum_{\vect{k},\sigma,\sigma^\prime}\left<\hat{\psi}^\dagger_{\alpha,\vect{k},\sigma}\vect{\sigma}_{\sigma\sigma^\prime}\hat{\psi}^{}_{\bar{\alpha},\vect{k},\sigma^\prime}\right> \mathcomma
\end{equation}
$s^{+-}$~superconductivity (SC) with order parameter 
\begin{equation}
 \Delta = \sum_{\vect{k}}\left<\hat{\psi}_{1,\vect{k},\uparrow}^\dagger\hat{\psi}_{1,-\vect{k},\downarrow}^\dagger-\hat{\psi}^\dagger_{2,\vect{k},\uparrow}\hat{\psi}^\dagger_{2,-\vect{k},\downarrow}\right>\mathcomma
\end{equation}
and charge density waves (iCDW) with an imaginary order parameter 
\begin{equation}
 \rho = -\frac{\mathrm{i}}{2}\sum_{\vect{k},\sigma}\left<\hat{\psi}_{1,\vect{k},\sigma}^\dagger \hat{\psi}_{2,\vect{k},\sigma}^{}-\hat{\psi}_{2,\vect{k},\sigma}^\dagger\hat{\psi}_{1,\vect{k},\sigma}^{}\right>
\end{equation}
associated with orbital magnetism. Thus, at this fixed point, the free energy~$F$ is a function of a combined order parameter, $F=F(\vect{M}^2+|\Delta|^2+\rho^2)$. 

Since iron-based superconductors are only close to this SO(6)-symmetric fixed point, the SDW or SC instabilities occur first, and iCDW~order has not been observed in any iron-based superconductor so far, although being close in energy. It is, however, a conceivable scenario that such order could nucleate around impurities in these materials, similar to SDW~order\cite{Millis.Morr.Schmalian.2001,GastiasoroAndersen2013,Gastiasoro2013-Dimers}. 
Such iCDW-type impurities break time-reversal symmetry and thereby are responsible for orbital magnetism. Thus we consider such iCDW~impurities as an example to demonstrate the emergence of a nontrivial phase in the impurity line in iron-based superconductors. 
Since $\rho$ is an Ising order parameter, it can nucleate with either sign around a given impurity site. 

An iCDW-type impurity at site $\vect{R}_i$ is described by 
\begin{align}
  \hat{U}_{\vect{R}_i}&=-\mathrm{i}\frac{U_0}{2}\sum_{s,\sigma}\mathrm{e}^{\mathrm{i}\vect{Q}\cdot(\vect{R}_i+\vect{R}_s)}  \nonumber \\ 
  & \qquad \times \Big[\hpsi_{1,i+s,\sigma}^\dagger\hpsi_{2,i+s,\sigma}^{} 
  -\hpsi_{2,i+s,\sigma}^\dagger\hpsi_{1,i+s,\sigma}^{}\Big]\mathcomma
\end{align}
and therefore each impurity breaks time-reversal symmetry. Here, $U_0$ is proportional to the iCDW order parameter~$\rho$ and an appropriate electron-electron interaction matrix element. 
Furthermore, such an impurity is associated with an orbital loop current which can be described by an Ising order parameter. 
For short-ranged impurities, the sum over lattice sites $s$ can, for example, be restricted to nearest neighbors~(NN). 
In momentum space, the corresponding matrix element is given by 
\begin{align}
  \left<\vect{k},\alpha\right|\hat{U}_{\vect{R}_i}\left|\vect{k}^\prime,\beta\right>&=-\mathrm{i}\frac{U_0}{2}\mathrm{e}^{\mathrm{i}(\vect{k}-\vect{k}^\prime)\cdot\vect{R}_i}\sum_s\mathrm{e}^{\mathrm{i}(\vect{k}-\vect{k}^\prime+\vect{Q})\cdot\vect{R}_s} \nonumber\\
  &\qquad \times \left[\updelta_{\alpha,1}\updelta_{\beta,2}-\updelta_{\alpha,2}\updelta_{\beta,1}\right]\mathcomma
\end{align}
and thus the scattering rate is given by 
\begin{align}
  \Gamma_{\alpha\beta\gamma\delta}(\vect{k}_1^{},\vect{k}_1^\prime,\vect{k}_2^{},\vect{k}_2^\prime)&=-\frac{ n_\mathrm{imp}U_0^2}{4} \sum_s\mathrm{e}^{\mathrm{i}(\vect{k}_1^{}-\vect{k}_1^\prime+\vect{Q})\cdot\vect{R}_s}\nonumber\\
  & \quad\times\sum_t\mathrm{e}^{\mathrm{i}(\vect{k}_2^{}-\vect{k}_2^\prime+\vect{Q})\cdot\vect{R}_t}\updelta_\mathrm{band}\mathperiod
\end{align}
Hence, even if impurities at different sites were to exhibit a different sign of the loop-current Ising order parameter, the scattering rate would be unaffected because it only depends on the order parameter squared. 
For the interband scattering process corresponding to the exchange of two electrons between the bands, $\alpha=\delta\neq\gamma=\beta$, it holds that $\updelta_\mathrm{band}=-1$. The interband scattering process from which a phase in the impurity line might arise is associated with $\delta_\mathrm{band}=+1$ and corresponds to a Cooper pair being scattered to the other band, that is, $\alpha=\gamma\neq\beta=\delta$. All other combinations of band indices yield $\delta_\mathrm{band}=0$, reflecting that this particular type  of impurities can only cause certain interband scattering processes. 

Keeping in mind that a global prefactor of $-1$ corresponds to a phase of $\pi$, we evaluate the imaginary part of the impurity line that might yield arbitrary phases. It is determined from the phase factors, 
\begin{align}
 &\Im \Gamma_{\alpha\beta\gamma\delta}(\vect{k}_1^{},\vect{k}_1^\prime,\vect{k}_2^{},\vect{k}_2^\prime) \nonumber \\ 
 &\quad \propto\Im \left[ \sum_s\mathrm{e}^{\mathrm{i}(\vect{k}_1^{}-\vect{k}_1^\prime+\vect{Q})\cdot\vect{R}_s}\sum_t\mathrm{e}^{\mathrm{i}(\vect{k}_2^{}-\vect{k}_2^\prime+\vect{Q})\cdot\vect{R}_t}\right] \nonumber \\
 &\quad = \sum_{s,t}\sin((\vect{k}_1^{}-\vect{k}_1^\prime+\vect{Q})\cdot\vect{R}_s+(\vect{k}_2^{}-\vect{k}_2^\prime+\vect{Q})\cdot\vect{R}_t)
\end{align}
which can be evaluated assuming a lattice possessing certain symmetries and a finite range of the impurity. As long as inversion symmetry is present in the crystal, the imaginary part of the impurity line is zero. However, phases of $0$ and $\pi$ are possible even in case of an inversion-symmetric lattice. For example, in case of zero incoming momenta, $\vect{k}_1=\vect{k}_2=\vect{0}$, and outgoing momenta $\vect{Q}$, $\vect{k}_1^\prime=\vect{k}_2^\prime=\vect{Q}$, an inversion-symmetric lattice, and short-ranged impurities that only affect neighboring sites, we find a phase of $\pi$ since  
\begin{equation}
  \Gamma_{1212}=-\frac{n_\mathrm{imp}}{4}(N_\mathrm{NN})^2 \mathcomma
\end{equation}
where $N_\mathrm{NN}$ is the number of nearest-neighbor sites. 
When, additionally, the lattice breaks inversion symmetry, even arbitrary phases are conceivable, also leading to suppression of $T_\mathrm{c}$, but with a different functional behavior. 
\section{Conclusion}\label{sec:conclusion}
We consider a two-band superconductor in the presence of impurities. Depending on the interaction leading to superconductivity, this model describes conventional or unconventional superconductivity which is known to react differently to the presence of impurities, also depending on whether the impurities are sensitive to the spin of the scattered electrons or not. 
Extended potential impurities, although insensitive to spin, can still break time-reversal symmetry, and in this paper we consider the effect of such impurities associated with orbital magnetism on the transition temperature. One example for the occurrence of this effect could be a competing state of order nucleated by the impurity. Such a scenario is conceivable in the case of iron-based superconductors, where imaginary charge density waves are a hidden state of order competing with superconductivity. 

Orbital magnetism, that as competing ordered state nucleates near impurities, manifests itself in a nontrivial phase in the impurity line of one interband scattering process, and we classify different limits by this phase. Our results for the transition temperature are summarized in Fig.~\ref{fig:suppression}. The trivial phase $\phi=0$ corresponds to the well-known situation: The transition temperature~$T_\mathrm{c}$ of conventional superconductors remains unaffected by impurities, whereas for unconventional superconductors, $T_\mathrm{c}$ is suppressed with increasing interband scattering rate, and even vanishes completely at a critical scattering rate. The functional behavior of $T_\mathrm{c}$ on the interband scattering rate corresponds to the functional behavior originally only associated with paramagnetic impurities by Abrikosov and Gorkov. For a phase of $\phi=\pi$, however, we find the reversed situation. Then, impurities are pair breaking for conventional superconductors with the same 
functional behavior, and there exists an analog of the 
Anderson theorem for unconventional superconductors. 
This scenario is indeed realized in case of the imaginary charge density wave state discussed in Refs.~\onlinecite{Chubukov2008,Podolsky2009,KangTesanovic2011}. 

Since the phase~$\phi$ in the impurity line is only defined relative to a similar phase in the BCS coupling matrix element~$V^{\alpha\bar{\alpha}}_{\vect{k},\vect{k}^\prime}$, this result can also be understood in terms of a redefinition of the electron operators in order to absorb the phase of the impurity line associated with the scattering process with rate $\Gamma_{1212}$, 
\begin{align}
\begin{split}
  \hpsi_{1,\vect{k},\sigma}&\rightarrow\hpsi^\prime_{1,\vect{k},\sigma}=\mathrm{e}^{\mathrm{i}\frac{\phi}{2}}\hpsi_{1,\vect{k},\sigma} \mathcomma\\
  \hpsi_{2,\vect{k},\sigma}&\rightarrow\hpsi^\prime_{2,\vect{k},\sigma}=\hpsi_{2,\vect{k},\sigma}\mathperiod
\end{split}
  \end{align}
This leaves the intraband scattering processes as well as the interband scattering process associated with rate~$\Gamma_{1221}$ unaffected, but entails a simultaneous rescaling of the BCS~coupling matrix element $V\rightarrow V^\prime=\mathrm{e}^{-\mathrm{i}\phi}V$. In the case of $\phi=\pi$ this corresponds to $V\rightarrow V^\prime=-V$, and thus, an attractive interaction under this transformation effectively becoming repulsive, and vice versa. Therefore, for a phase of $\phi=\pi$ we find an Anderson theorem for the $s^{+-}$~pairing state, whereas the transition temperature of the $s^{++}$~pairing state is suppressed according to the Abrikosov-Gorkov law. 

In the intermediate regime, impurities are pair breaking for both pairing states, but there is no critical interband scattering rate at which superconductivity is suppressed completely. As an example, we consider $\phi=\frac{\pi}{2}$, and find linear suppression of $T_\mathrm{c}$ for small interband scattering rates, and exponential suppression of $T_\mathrm{c}$ in the dirty limit. 

In conclusion, in the presence of impurities associated with orbital magnetism, pair breaking due to interband scattering does not only occur in unconventional superconductors, and the robustness of $T_\mathrm{c}$ against impurities does not necessarily imply conventional superconductivity. 

Additionally, we give a condition under which spin density waves as they occur in iron-based superconductors are also protected against impurities. We find that spin density waves are stable against the impurities associated with orbital magnetism that we considered as an example in Sec.~\ref{example}, but prone to intraband scattering breaking particle-hole symmetry.
Thus, we expect no change of the SDW and $s^{+-}$~SC transition temperatures in the case of iCDW impurities. However, particle-hole symmetry-breaking intraband scattering will suppress SDW order while leaving SC order unchanged such that in a coexisting state of spin density wave order and superconductivity $T_\mathrm{c}$ may increase as demonstrated in Refs.~\onlinecite{FernandesVavilovChubukov2012,LiXu2010}. 

We note that the effect of {\it spin-magnetic} impurities (not considered here microscopically) on the superconductive transition has been addressed recently in Ref.~\onlinecite{Korshunov2014}. 
Their results are consistent with our general  symmetry analysis of Sec.~\ref{sec:symmetryanalysis}, while our diagrammatic calculation of Secs.~\ref{sec:averaging} and \ref{sec:tc} focuses on the other case of orbital-magnetic impurities and a possible microscopic mechanism for such impurities.
\section*{Acknowledgments}
We thank P.~J.~Hirschfeld and A.~Shnirman for helpful discussions. This work was supported by the Deutsche Forschungsgemeinschaft through DFG-SPP~1458 `Hochtemperatursupraleitung in Eisenpniktiden'.
The work of SVS was partially supported
by the Alexander von Humboldt Foundation
through the Feodor Lynen Research Fellowship and by the
NSF grants DMR-1001240, DMR-1205303, and PHY-1125844.
\appendix
\section{Derivation of the stability condition for density wave phases in the presence of disorder} \label{app:anderson-dw}
This Appendix is devoted to the proof of the condition~\eqref{AndersonDensityWave} for stability of SDW order, $\{\hat{W},\hat{O}\} = 0$. 
Let us first consider two Hermitian matrices $A$ and $B$ with (real) eigenvalues $\{\lambda_A^{(i)}\}$ and $\{\lambda_B^{(i)}\}$, respectively. Furthermore, let us denote the eigenvalues of $A+B$ by $\{\lambda_{A+B}^{(i)}\}$. When $A$ and $B$ anticommute, it holds
\begin{equation}
 (A+B)^2 = A^2 + B^2 + \{A,B\} = A^2 + B^2\mathperiod
\end{equation} 
Since $A^2$ and $B^2$ commute, they can be diagonalized simultaneously and, thus, we have
\begin{equation}
 \left(\lambda_{A+B}^{(i)}\right)^2 = \left(\lambda_{A}^{(i)}\right)^2 +\left(\lambda_{B}^{(\pi(i))}\right)^2\mathcomma
\end{equation} 
with some permutation $\pi$. In particular, this implies
\begin{equation}
 \min_i \left|\lambda_{A+B}^{(i)}\right| \geq \min_{i}\left|\lambda_{A}^{(i)}\right|,\, \min_{i}\left|\lambda_{B}^{(i)}\right|\mathperiod \label{MinimumSpectr}
\end{equation}

When condition~\eqref{AndersonDensityWave} is satisfied, it holds that 
\begin{align}
\begin{split}
 \left\{\begin{pmatrix} \hat{\xi}_1 & 0 \\ 0 & -\hat{\xi}_1 \end{pmatrix} 
 + \hat{W} ,\hat{O}\right\} = \\ \left\{\begin{pmatrix} \hat{\xi}_1 & 0 \\ 0 & -\hat{\xi}_1 \end{pmatrix},\begin{pmatrix} 0 & m \\ m^\dagger & 0 \end{pmatrix}\right\} +  \left\{\hat{W},\hat{O}\right\} = 0\mathperiod
\end{split}
 \end{align} 
Due to \eqref{MinimumSpectr}, the gap cannot be reduced by $\hat{W}$ and, consequently, the density wave is stable against any disorder configuration that anticommutes with its order parameter. 
\section{Calculation of the generalized Cooperon ladder}
\label{app:cooperon-ladder}
This Appendix provides details on the calculation of the generalized form of the Cooperon ladder, denoted by~$C_{\alpha}$. A single rung of the ladder~$C_\alpha$ is given by 
\begin{widetext}
\begin{align}
 & \vcenter{\hbox{\includegraphics[height=5em]{./cooperon1_rev}}} +\vcenter{\hbox{\includegraphics[height=5em]{./cooperon2a_rev}}}\times\left(1+\vcenter{\hbox{\includegraphics[height=5em]{./cooperon3a_rev}}}+\vcenter{\hbox{\includegraphics[height=5em]{./cooperon3b_rev}}}+\cdots\right)\times\vcenter{\hbox{\includegraphics[height=5em]{./cooperon2b_rev}}} \nonumber \\ 
  &\quad = \Gamma_\alpha{\int_{\vect{k}}}G_{\alpha,\vect{k}}(\nu_n)G_{\alpha,-\vect{k}}(-\nu_n)\left[1 + \frac{\Gamma_{12}^2}{\Gamma_{\alpha}}{\int_{\vect{k}^\prime}} G_{\bar{\alpha},\vect{k}^\prime}(\nu_n)G_{\bar{\alpha},-\vect{k}^\prime}(-\nu_n) \sum_{m=0}^\infty\left(\Gamma_{\bar{\alpha}}{\int_{\vect{k}^{\prime\prime}}}G_{\bar{\alpha},\vect{k}^{\prime\prime}}(\nu_n)G_{\bar{\alpha},-\vect{k}^{\prime\prime}}(-\nu_n)\right)^m\right] \nonumber \\
  &\quad = \frac{\pi\rho_\mathrm{F}\Gamma_\alpha|\nu_n|+(\pi\rho_\mathrm{F})^2\Gamma_{12}(\Gamma_\alpha+\Gamma_{12})}{(\pi\rho_\mathrm{F}\Gamma_{12}+|\nu_n|)[\pi\rho_\mathrm{F}(\Gamma_\alpha+\Gamma_{12})+|\nu_n|]}
  \label{eq:singlerung}
\end{align}
\end{widetext}
where the second term appears in addition to the usual Cooperon ladder for scattering in single-band models or in models with intraband scattering only. 
In Eq.~\eqref{eq:singlerung}, the propagators drawn in light gray are only shown for clarification of the respective scattering processes and not part of the calculation. The last line has been obtained by performing the energy integration. 

In order to obtain the full generalized Cooperon ladder, the result for a single rung, Eq.~\eqref{eq:singlerung}, is summed, yielding 
\begin{align}
  C_{\alpha}(\nu_n)&=\sum_{m=0}^\infty\left(\frac{\pi\rho_\mathrm{F}\Gamma_\alpha|\nu_n|+(\pi\rho_\mathrm{F})^2\Gamma_{12}(\Gamma_\alpha+\Gamma_{12})}{(\pi\rho_\mathrm{F}\Gamma_{12}+|\nu_n|)[\pi\rho_\mathrm{F}(\Gamma_\alpha+\Gamma_{12})+|\nu_n|]}\right)^m \nonumber\\
  &=\frac{(\pi\rho_\mathrm{F}\Gamma_{12}+|\nu_n|)[\pi\rho_\mathrm{F}(\Gamma_{\alpha}+\Gamma_{12})+|\nu_n|]}{|\nu_n|(2\pi\rho_\mathrm{F}\Gamma_{12}+|\nu_n|)}\mathperiod
\end{align}

\end{document}